\documentclass[12pt]{article}
\pdfoutput=1

\usepackage[utf8]{inputenc}
\usepackage{jheppub}
\usepackage{epsfig}
\usepackage{bbm}
\usepackage{bbding}
\usepackage{amsmath}
\usepackage{amsfonts}
\usepackage{amssymb}
\usepackage{mathtools}
\usepackage{dsfont}
\usepackage{bm}
\usepackage{graphicx}
\usepackage{longtable}
\usepackage{pdflscape}
\usepackage{subcaption}
\usepackage{color}
\usepackage{psfrag}
\usepackage[capitalise]{cleveref}
\usepackage{hyperref}
\usepackage{tikz}
\usetikzlibrary{positioning,arrows.meta}
\usepackage{pgfplots}
\usepackage{pgfplotstable}
\usepackage{subfiles}
\usepackage{longtable}
\usepackage{tabularx}
\usepackage{ltablex}
\usepackage{booktabs}
\usepackage[export]{adjustbox}
\usepackage{listings}
\usepackage{multirow} 
\usepackage{cancel,slashed}
\usepackage{bbm}
\usepackage{graphicx}
\usepackage{latexsym}
\usepackage{color}
\usepackage{transparent}
\usepackage{mathtools}
\usepackage{array}
\usepackage{makecell}
\usepackage{graphics,psfrag}
\usepackage{placeins}
\usepackage{nowidow}
\usepackage{listings}
\usepackage[normalem]{ulem}
\usepackage{environ}

\pgfplotsset{
        compat=1.9,
        compat/bar nodes=1.8,
    }

\definecolor{bluetto}{HTML}{0088ff}
\definecolor{snowfl}{HTML}{00ace6}

\makeatletter
\def\@xfootnote[#1]{%
	\protected@xdef\@thefnmark{#1}%
	\@footnotemark\@footnotetext}
\makeatother

\definecolor{primcol}{HTML}{1e6091}
\definecolor{seccol}{HTML}{52b69a}
\definecolor{tercol}{HTML}{b5e48c}
\definecolor{redcol}{HTML}{9a031e}

\DeclareMathOperator{\SU}{SU}
\DeclareMathOperator{\SO}{SO}

\newcommand{\de}{\partial}

\newcommand{\PP}{\mathbb{P}}

\newcommand{\ZZ}{\mathbb{Z}}

\newcommand{\coma}{\, , \quad}
\newcommand{\fstop}{\, .}

\newcommand{\UV}{\text{UV}}
\newcommand{\IR}{\text{IR}}

\renewcommand{\epsilon}{\varepsilon}

\makeatletter
\newsavebox{\measure@tikzpicture}
\NewEnviron{scaletikzpicturetowidth}[1]{%
  \def\tikz@width{#1}%
  \def\tikzscale{1}\begin{lrbox}{\measure@tikzpicture}%
  \BODY
  \end{lrbox}%
  \pgfmathparse{#1/\wd\measure@tikzpicture}%
  \edef\tikzscale{\pgfmathresult}%
  \BODY
}
\makeatother

\def\IX{{\bf {X}}}

\def\im{{\rm Im \,}}
\def\re{{\rm Re \,}}

\catcode`\@=11   
\newdimen\@rotdimen
\newbox\@rotbox  

\def\@vspec#1{\special{ps:#1}}
\def\@rotstart#1{\@vspec{gsave currentpoint currentpoint translate
		#1 neg exch neg exch translate}}
\def\@rotfinish{\@vspec{currentpoint grestore moveto}}
\def\@rotr#1{\@rotdimen=\ht#1\advance\@rotdimen by\dp#1%
	\hbox to\@rotdimen{\hskip\ht#1\vbox to\wd#1{\@rotstart{90 rotate}%
			\box#1\vss}\hss}\@rotfinish}
\def\@rotl#1{\@rotdimen=\ht#1\advance\@rotdimen by\dp#1%
	\hbox to\@rotdimen{\vbox to\wd#1{\vskip\wd#1\@rotstart{270 rotate}%
			\box#1\vss}\hss}\@rotfinish}%
\def\@rotu#1{\@rotdimen=\ht#1\advance\@rotdimen by\dp#1%
	\hbox to\wd#1{\hskip\wd#1\vbox to\@rotdimen{\vskip\@rotdimen
			\@rotstart{-1 dup scale}\box#1\vss}\hss}\@rotfinish}%
\def\@rotf#1{\hbox to\wd#1{\hskip\wd#1\@rotstart{-1 1 scale}%
		\box#1\hss}\@rotfinish}%
\def\rotate{\@ifnextchar[{\@rotate}{\@rotate[l]}}
\def\@rotate[#1]#2{\setbox\@rotbox=\hbox{#2}\@nameuse{@rot#1}\@rotbox}

\catcode`\@=12

\tikzset{
    partial ellipse/.style args={#1:#2:#3}{
        insert path={+ (#1:#3) arc (#1:#2:#3)}
    }
}

\hypersetup{
	pdftitle={Thraxions: Towards Full String Models},    
	pdfauthor={\textcopyright\ Federico Carta, Alessandro Mininno, Nicole Righi, Alexander Westphal},     
	pdfsubject={HEP},   
	pdfcreator={pdfLaTex},   
	pdfproducer={LaTex}, 
	pdfkeywords={},
}

\allowdisplaybreaks[1] 

\crefname{figure}{Figure}{Figures}
\crefname{table}{Table}{Tables}

\begin{document}
	\pagestyle{plain}

	\makeatletter
	\@addtoreset{equation}{section}
	\makeatother
	\renewcommand{\theequation}{\thesection.\arabic{equation}}
	\pagestyle{empty}
\rightline{IFT-UAM/CSIC-21-106}
\rightline{ZMP-HH/21-18}
\rightline{DESY 21-152}
\vspace{1.5cm}

\begin{center}
	\LARGE{\bf Thraxions: Towards Full String Models}\\
	\large{Federico Carta,\textsuperscript{1} Alessandro Mininno,\textsuperscript{2,3} Nicole Righi,\textsuperscript{4} \\Alexander Westphal\textsuperscript{4}\\[4mm]}
	\footnotesize{\textsuperscript{1}Department of Mathematical Sciences, Durham University,\\
	Durham, DH$1$ $3$LE, United Kingdom\\
		\textsuperscript{2}Instituto de F\'{\i}sica Te\'orica IFT-UAM/CSIC,\\
		C/ Nicol\'as Cabrera 13-15, 
		Campus de Cantoblanco, 28049 Madrid, Spain\\
	\textsuperscript{3}II. Institut f\"ur Theoretische Physik, Universit\"at Hamburg,\\
Luruper Chaussee 149, 22607 Hamburg, Germany\\
	\textsuperscript{4}Deutches Electronen-Synchrotron, DESY,\\ Notkestra\ss e 85, 22607 Hamburg, Germany}\\
	\footnotesize{\href{mailto:federico.carta@durham.ac.uk}{federico.carta@durham.ac.uk}, \href{mailto:alessandro.mininno@desy.de}{alessandro.mininno@desy.de}, \href{mailto:nicole.righi@desy.de}{nicole.righi@desy.de}, \href{mailto:alexander.westphal@desy.de}{alexander.westphal@desy.de}}

\vspace*{4mm}

\small{\bf Abstract} 
\\[4mm]
\end{center}
\begin{center}
\begin{minipage}[h]{\textwidth}
We elucidate various aspects of the physics of thraxions, ultra-light axions arising at Klebanov-Strassler multi-throats in the compactification space of IIB superstring theory. We study the combined stabilization of K\"ahler moduli and thraxions, showing that under reasonable assumptions, one can solve the combined problem both in a KKLT and a LVS setup. We find that for non-minimal multi-throats, the thraxion mass squared is three-times suppressed by the throat warp factor. However, the minimal case of a double-throat can preserve the six-times suppression as originally found. We also discuss the backreaction of a non-vanishing thraxion vacuum expectation value on the geometry, showing that it induces a breaking of the imaginary self-duality condition for $3$-form fluxes. This in turn breaks the Calabi-Yau structure to a complex manifold one. Finally, we extensively search for global models which can accommodate the presence of multiple thraxions within the database of Complete Intersection Calabi-Yau orientifolds. We find that each multi-throat system holds a single thraxion. We further point out difficulties in constructing a full-fledged global model, due to the generic presence of frozen-conifold singularities in a Calabi-Yau orientifold. For this reason, we propose a new \href{https://www.desy.de/~westphal/orientifold_webpage/cicy_thraxions.html}{database} of CICY orientifolds that do not have frozen conifolds but that admit thraxions.
\end{minipage}
\end{center}

	\newpage
	\setcounter{page}{1}
	\pagestyle{plain}
	\renewcommand{\thefootnote}{\arabic{footnote}}
	\setcounter{footnote}{0}
	
	\tableofcontents
	
	
\section{Introduction}

A generic prediction of string phenomenology models is the presence of axion-like particles in the four-dimensional effective theory. Often arising in the compactification process as integrals of higher $p$-form gauge potentials on $p$-cycles of the internal space, the number and properties of these particles are determined essentially by the ten-dimensional origin.

Some unique properties of string axions make them interesting to study from a phenomenological point of view~\cite{Banks:2003sx,Svrcek:2006yi,Grimm:2007hs,Arvanitaki:2009fg,Cicoli:2012sz}, for example in the context of inflation, or as possible dark matter constituents. Moreover, their features are of central prominence in the Swampland program~\cite{Vafa:2005ui,Ooguri:2006in}, especially in the context of the Weak Gravity Conjecture~\cite{Arkani-Hamed:2006emk}. Hence, axions represent a promising class of particles which could provide information about the underlying theory of quantum gravity at the level of testable physics. Many of the next-generation experiments will partially cover the parametric space where string axions are expected to live. In addition, thanks to the discovery of gravitational waves~\cite{PhysRevLett.116.061102}, there is now a completely new window where one could detect their effect on gravitational phenomena, such as the superradiance instability of binary black holes (BHs)~\cite{Arvanitaki:2009fg} (for recent progress in linking large-scale CY database scans with the BH superradiance constraints for axions see e.g.~\cite{Mehta:2021pwf}).

In this paper, we focus on type IIB superstring theory compactified to $4$d on a compact six dimensional  Calabi-Yau (CY) orientifold with 3-form fluxes~\cite{Giddings:2001yu,Grimm:2004uq}. At low energy and large volume, the supergravity approximation holds. In order to extract phenomenological properties of these axionic particles (such as their mass and decay constant), we should then study the moduli stabilization problem in the context of the low-energy effective $4$d $\mathcal{N}=1$ supergravity. Lots of efforts have been spent in this direction since two different moduli stabilization procedures for type IIB on CYs were proposed by Kachru, Kallosh, Linde and Trivedi (dubbed as KKLT~\cite{Kachru:2003aw}) and subsequently by Balasubramanian, Berglund, Conlon and Quevedo (whose model goes under the name of Large Volume Scenario, or LVS~\cite{Balasubramanian:2005zx}). However, the stabilization of one particular class of axions is still left unexplored.

\emph{Thraxions}, or throat-axions~\cite{Hebecker:2018yxs}, are a recently discovered class of ultra-light axionic modes arising whenever the CY admits a system of multiple warped throats (multi-throat) sharing some common 3-cycle $\mathcal{B}$, near a conifold point in complex structure moduli space. In such a case, it is in fact possible to reduce the $2$-form RR and NS potentials $C_2$ and $B_2$ on the family of sectional $S^2\subset \mathcal{B}$, generating new axions as the lowest-lying radial Kaluza-Klein (KK) mode in the low energy effective theory. These axions were found to be parametrically lighter than any other particle in the spectrum. Their mass squared is suppressed by six times the warp factor $\omega_{\IR}$ of the throats, while the warped-throat KK modes of, e.g. the throat complex-structure (c.s.) modulus, receive a double suppression only. Since the thraxions own such unique features, it is important to explore their behavior in a fully stabilized setup in order to connect them with axion phenomenology. 

The aim of this work is to discuss two relevant questions that were left behind in the original paper~\cite{Hebecker:2018yxs}. First, we study the effect of a non-vanishing thraxion vacuum expectation value (VEV) at the level of the $\SU(3)$-structure torsion classes of the compactification space.
We find that for a non-vanishing VEV, the compactification space fails to be CY and becomes simply a complex manifold. We understand this as a breakdown of the imaginary self-dual condition (ISD) of the $G_3$-flux, and we relate quantitatively the size of the thraxion VEV with the size of the ISD breaking. 
Second, we study the interplay between thraxions and K\"ahler moduli stabilization. 

In particular, we find that in general the thraxion potential receives potentially non-vanishing corrections which lift the mass squared to $\sim \omega_{\IR}^3$ only. After explaining why this is the case, we show that these cross terms in general do not vanish in multi-throats consisting of at least three joined throats. Conversely, in the simplest class of double-throats, avoiding the cross terms reduces to essentially a single concretely realizable condition on the periods of double-throats.

Among the necessary conditions for the absence of these corrections on generic multi-throats, we find that each multi-throat system must admit one single thraxion. Motivated by this, we perform a quantitative and systematic search for a realistic global model containing thraxions within the Complete Intersection Calabi-Yau (CICY) orientifold database introduced in~\cite{Carta:2020ohw}. Remarkably, we find that in all CICY orientifolds with O$3$/O$7$-planes which allow for the presence of thraxions there is always a single thraxion per multi-throat system, and multiple multi-throat systems are also realized. We compile a database of all the CICY orientifolds supporting thraxions, which extends the one given in~\cite{Carta:2020ohw}.

This paper is organized as follows. In Section~\ref{sec:reviewThraxions} we review the r\^ole of thraxions in type IIB CY compactifications with fluxes. In Section~\ref{sec:ThraxionsTorclass} we show how the backreaction of thraxions on the geometry breaks the conformal CY condition of GKP-like solutions~\cite{Giddings:2001yu} using the torsion classes defined on the $6$d internal manifold used in the compactification. In Section~\ref{sec:ModuliStabilization} we show that K\"ahler moduli stabilization induces corrections to the thraxion potential. We discuss this issue both in KKLT and in LVS setups. In Section~\ref{sec:MultiThrax-database} we introduce a new database constructed from the CICY orientifolds one proposed in~\cite{Carta:2020ohw}. This database can be found at this \href{https://www.desy.de/~westphal/orientifold_webpage/cicy_thraxions.html}{link} and contains all the orientifolds of the CICYs that do not have frozen conifolds (in the sense of~\cite{Carta:2020ohw}) and in which there is at least one thraxion. For technical details, we provide two appendices. In Appendix~\ref{app:modstabdetails} we give detailed examples for the K\"ahler moduli stabilization following the approximations introduced in Section~\ref{sec:ModuliStabilization}. 

Finally, in Appendix~\ref{app:KKLT-repth} we show that the masses for the axion and saxion, $c$ and $b$, in the KKLT AdS vacuum are consistent with results from the representation theory of a Quantum Field Theory (QFT) defined in an AdS background. 

\section{Thraxions in Flux Compactification}
\label{sec:reviewThraxions}
Let us consider type IIB superstring theory compactified on a compact CY threefold $\IX$ whose volume is sufficiently larger than the string scale. The manifold $\IX$ has a c.s. moduli space $\mathcal{M}_{cs}$ and a K\"ahler moduli space $\mathcal{M}_{k}$ respectively of complex dimension $h^{2,1}$ and $h^{1,1}$. The low energy effective theory is $4$d $\mathcal{N}=2$ supergravity with a field content of $h^{1,1}$ hypermultiplets and $h^{2,1}$ vector multiplets.\footnote{There is also the gravity multiplet and the double-tensor multiplet. Both will play no role in this paper.}

Let $H_3(\IX, \mathbb{Z})$ be the third homology group of $\IX$ with integer coefficients. We fix an integral basis for $H_3(\IX, \mathbb{Z})$ consisting of $b_3=2h^{2,1}+2$ 3-cycles $\mathcal{A}_i$, $\mathcal{B}^j$, $i,j=1,\ldots h^{2,1}+1$. This integral basis is symplectic, meaning that the 3-surfaces intersect as
\begin{equation}
\mathcal{A}_i\cap \mathcal{B}^j=- \mathcal{B}^j\cap \mathcal{A}_i=\delta^{\ j}_i\coma \mathcal{A}_i\cap\mathcal{A}_j= \mathcal{B}^i\cap \mathcal{B}^j=0 \fstop
\end{equation}
The c.s. moduli space $\mathcal{M}_{cs}$ is special K\"ahler, and one can define special coordinates on it as follows:
\begin{equation}
z_i\coloneqq\int_{\mathcal{A}_i}\Omega\coma F^j\coloneqq\int_{\mathcal{B}^j}\Omega\coma 
\label{eq:spcoord}
\end{equation}
where $\Omega$ is the nowhere vanishing holomorphic 3-form that exists on $\IX$ since $\IX$ is CY. In particular, we take \eqref{eq:spcoord} as the definition of the c.s. moduli $z_i$.

Generic pointlike singularities of $\IX$ arise at specific codimension 1 loci in $\mathcal{M}_{cs}$, where one of the c.s. moduli $z_i$ vanishes. Such singularities are called conifold points~\cite{Candelas:1989js,Candelas:1989ug}. It is important to discuss now a crucial difference regarding conifold singularities in a compact CY, compared to a non-compact one. In a non-compact setting, it is possible to have a conifold singularity in which a single one of the c.s. moduli $z_i$ vanishes. From \eqref{eq:spcoord} this implies that a single $\mathcal{A}$-cycle vanishes. On the other hand, in a compact setting a conifold singularity is only possible if two or more $\mathcal{A}$-cycle related in homology shrink to zero volume~\cite{Candelas:1990rm}.\footnote{The reason for this arises from the subset of conifold singularities which admit a resolution phase. In such a case, the fact that a single 3-cycle shrinks to zero size will induce a breaking of the K\"ahler condition on the resolution side of the transition.} Throughout the paper, we will call this latter case \emph{multi-conifold}, to distinguish it with the non-compact case in which a single $\mathcal{A}$-cycle vanishes.

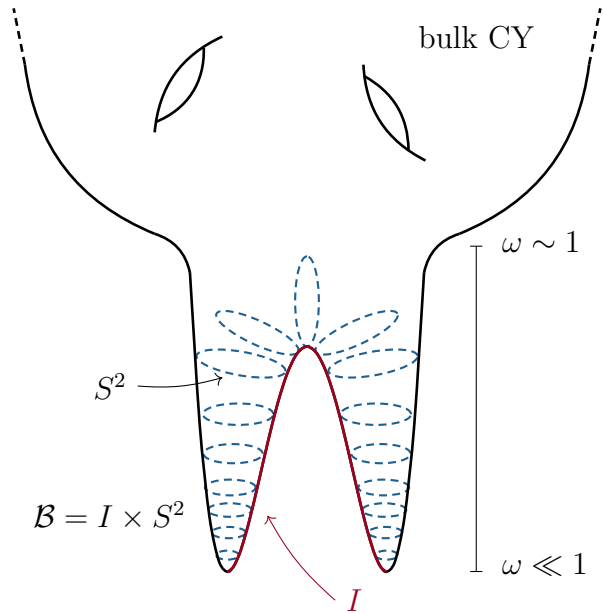
\begin{figure}[!htp]
    \centering
    \begin{tikzpicture}[scale=0.75]
    \draw[densely dashed,primcol,thick] (1.41,-3.7) ellipse (2mm and 0.8mm);
    \draw[densely dashed,primcol,thick] (1.39,-3.3) ellipse (3mm and 1mm);
    \draw[densely dashed,primcol,thick] (1.37,-2.9) ellipse (3.9mm and 1.2mm);
    \draw[densely dashed,primcol,thick] (1.36,-2.5) ellipse (4.5mm and 1.4mm);
    \draw[densely dashed,primcol,thick] (1.31,-1.9) ellipse (5.25mm and 1.7mm);
    \draw[densely dashed,primcol,thick] (1.22,-1.2) ellipse (6.25mm and 1.9mm);
    \draw[rotate=10,densely dashed,primcol,thick] (1.1,-0.5) ellipse (8mm and 2.1mm);
    \draw[rotate=25,densely dashed,primcol,thick] (0.9,-0.15) ellipse (8mm and 2.1mm);
    \draw[rotate=90,densely dashed,primcol,thick] (0.8,0) ellipse (8mm and 2.1mm);
    \draw[densely dashed,primcol,thick] (-1.41,-3.7) ellipse (2mm and 0.8mm);
    \draw[densely dashed,primcol,thick] (-1.39,-3.3) ellipse (3mm and 1mm);
    \draw[densely dashed,primcol,thick] (-1.37,-2.9) ellipse (3.9mm and 1.2mm);
    \draw[densely dashed,primcol,thick] (-1.36,-2.5) ellipse (4.5mm and 1.4mm);
    \draw[densely dashed,primcol,thick] (-1.31,-1.9) ellipse (5.25mm and 1.7mm);
    \draw[densely dashed,primcol,thick] (-1.22,-1.2) ellipse (6.25mm and 1.9mm);
    \draw[rotate=-10,densely dashed,primcol,thick] (-1.1,-0.5) ellipse (8mm and 2.1mm);
    \draw[rotate=-25,densely dashed,primcol,thick] (-0.9,-0.15) ellipse (8mm and 2.1mm);
    \draw[scale=1, domain=-2.075:2.075, smooth, variable=\x, line width=1pt,black] plot ({\x}, {\x*\x*\x*\x-4*\x*\x});
    \draw[scale=1, domain=-1.414:1.414, smooth, variable=\x, line width=1.05pt,redcol] plot ({\x}, {\x*\x*\x*\x-4*\x*\x});
    \draw[line width=1pt] (-2.075,1.31)to[bend right] (-2.7,2);
    \draw[line width=1pt] (2.075,1.31)to[bend left] (2.7,2);
    \draw[line width=1pt] (2.7,2) to[bend right] (5,5);
    \draw[line width=1pt] (-2.7,2) to[bend left] (-5,5);
    \draw[line width=1pt,densely dashed] (5,5) to[bend right=0] (5.2,6);
    \draw[line width=1pt,densely dashed] (-5,5) to[bend left=0] (-5.2,6);
    \draw[line width=1pt] (1.01,4.8) to[bend left=30] (1.8,3.48);
    \draw[line width=1pt] (2.1,3.3) to[bend left] (1,5);
    \draw[line width=1pt] (-1.83,5.3) to[bend left=35] (-2.69,3.98);
    \draw[line width=1pt] (-2.7,3.8) to[bend left] (-1.5,5.5);
    \node at (3,5.5) {bulk CY};
    \draw[thin,|-|] (3,-4) -- node[pos=0,right,xshift=0.2cm,yshift=0.1cm] {$\omega\ll 1$} node[pos=1, right,xshift=0.2cm,yshift=0.05cm] {$\omega\sim 1$} (3,1.79);
    \draw[thin,->] (-3,-0.7) to[bend right=10] node[pos=0,left] {$S^2$} (-1.5,-0.55);
    \node at (-3.5,-3) {$\mathcal{B}=I\times S^2$};
    \draw[thin,->,redcol] (0.5,-4.5) to[bend left=10] node[pos=0,right] {$I$} (-0.7,-3);
    \end{tikzpicture}
    \caption{Example of a double-throat system with one thraxion.}
    \label{fig:dt}
\end{figure}

Let us consider a multi-conifold singularity on $\IX$, in which a set of $n$ $\mathcal{A}$-cycles vanish, and they satisfy $m$ homology relations of the form 
\begin{equation}
p^I_i[\mathcal{A}]_i=0\fstop
\label{eq:homology}
\end{equation}
Eq. \eqref{eq:homology} only leaves $n-m$ $\mathcal{A}_i$-cycles independent. For each one of them, there is a symplectic dual $\mathcal{B}_i$-cycle in $H^3(\IX)$. Geometrically, this $\mathcal{B}_i$-cycle interpolates between $d_i$ singular points. The numerical value of $d_i$ can be determined as a function of the homology relations coefficients $p_i^I$, essentially determining which independent $\mathcal{B}_i$-cycle intersect which of the original $n$ $\mathcal{A}_i$-cycles.  We depict this schematically in Figure~\ref{fig:dt}, for the simplified case of $n=2, \ m=1$. Notice that in this picture, for ease of exposition, we draw two finite-size long throats rather than two conifold points. We will see later that this is the relevant setup once fluxes are turned on and the orientifold projection breaking to $\mathcal{N}=1$ is imposed. 

As it can be seen from Figure~\ref{fig:dt}, for $n=2, \ m=1$, the interpolating $\mathcal{B}$-cycle is, as a topological space, homeomorphic to $I\times S^2$, where $I$ is a finite size interval connecting the two singular points. The situation generalizes easily for $n,m >1$. In such a case the $n-m$ interpolating 3-cycles will be topologically homeomorphic to $Y^i_{d_i}\times S^2$ where $i=1,\ldots n-m$, and $Y_{d_i}$ is the complete graph with $d_i$ nodes.

We now \emph{define} the thraxion field as the dimensional reduction of the Ramond-Ramond (RR) 2-form gauge potential on the interpolating family of $S^2$ sphere discussed above, namely,
\begin{equation}
c\coloneqq \int_{S^2}  C_2 \fstop
\end{equation}
Hence, on the deformed side $c=c(r)$ does not constitute a true harmonic zero mode, but the lowest radial KK-mode in the multi-throat. Moreover, the decay constant can be computed from dimensional reduction of the $F_3\wedge \star F_3$ term over the sectional $S^2$ on which $c$ is defined. The result can be found in Appendix D, starting from Eqs. (108),(109) of ref. \cite{Hebecker:2018yxs} leading to Eq. (113) ibd. 

Given a conifold singularity, it is often also possible to perform a small resolution of it, producing extra 2-cycles.\footnote{This is not always the case. A famous example of a compact CY admitting a conifold singularity without resolution branch is the mirror quintic~\cite{Candelas:1990rm}.} Going from the deformed to the resolved phase is known as a conifold transition \cite{Candelas:1989fd}. Let us call $\tilde{\IX}$ the manifold on the resolution side. In a conifold transition, in a compact setting, the Hodge numbers change as $\tilde{h}^{1,1}>h^{1,1}$,  $\tilde{h}^{2,1}<h^{2,1}$. In particular, on the resolved side, there will be $\Delta h^{1,1}\coloneqq\tilde{h}^{1,1}-h^{1,1}$ extra resolution 2-cycles compared to the deformed side. This number of 2-cycles is equal to the number of the homology relations among the conifolds in the deformed side. On the resolved side, thraxions correspond to massless axions coming from the integrals of the 2-form over these $\Delta h^{1,1}$ independent resolutions $\mathbb{P}^1$. 

It is believed that the presence of conifold singularities and conifold transitions is very generic. It has been conjectured that all the CY manifolds are connected with each other by a web of conifold transitions~\cite{Reid1987,Candelas:1989fd}. It has also been shown that this statement holds true in numerous closed classes of examples~\cite{Chiang:1995hi,Avram:1997rs,Batyrev:1993oya,Batyrev:2008rp,Lynker:1995sy,Batyrev:1998kx}. Therefore, the existence of thraxions is a generic prediction of any IIB CY compactification, at the $\mathcal{N}=2$ level.

We now consider the introduction of an orientifold projection, and fluxes, in order to break supersymmetry to $\mathcal{N}=1$. For concreteness, we focus on the case in which the orientifold projection has O3/O7 fixed loci. The orientifold involution induces a splitting of the $H^{2,1}(\IX, \mathbb{Z})$ and $H^{1,1}(\IX,\mathbb{Z})$ cohomology groups into the direct sum of vector spaces $H^{2,1}(\IX, \mathbb{Z})\simeq H^{2,1}_+\oplus H^{2,1}_-$ and $H^{1,1}(\IX, \mathbb{Z})\simeq H^{1,1}_+\oplus H^{1,1}_-$. The dimensions $h_{1,1}^{\pm}$ (resp. $h_{2,1}^{\pm}$) of the eigenspaces $H^{1,1}_\pm$ (resp. $H^{2,1}_\pm$) fixes the number of fields present in the $4d \ \mathcal{N}=1$ effective theory~\cite{Grimm:2004uq}. These fields are the axiodilaton $\tau=C_0 +i e^{\phi}$,\footnote{Unless explicitly stated, we will set $C_0$ directly to zero in the following expressions.} $h^{2,1}_-$ c.s. moduli $z_i$, $h^{1,1}_-$ 2-form moduli $G^I$ and $h_{1,1}^+$ K\"ahler moduli $T_\alpha$. In particular, from the point of view of representation theory of the 4d $\mathcal{N}=1$ SUSY algebra, thraxions are the lowest component of a scalar chiral superfield $G^I$, $I=1,\dots,m$, i.e.
\begin{equation}
    G^I=\frac{1}{2 \pi \alpha' }\int_{\Sigma_I}(C_2-\tau B_2)=c^I-\tau b^I\fstop
    \label{eq:Gfieldsdef}
\end{equation}

Clearly, not all CY orientifolds  will support the presence of thraxions: for example many orientifolds are such that $h^{1,1}_-=0$. In order for thraxions to be present in a $\mathcal{N}=1$ setup, at least two crucial conditions must hold true:
\begin{itemize}
\item The orientifold projection must leave the conifold transition intact.
\item In the quotient space, a multi-conifold with interpolating $\mathcal{B}$-cycles must still exist.
\end{itemize}
It has been shown in~\cite{Carta:2020ohw} that orientifolds supporting thraxions exist within the set of complete intersection CYs. We will discuss at length these explicit models in Section~\ref{sec:MultiThrax-database}.

Having included an orientifold, for reasons of both tadpole cancellation and moduli stabilization, we will consider the addition of $3$-form fluxes. Flux quanta are defined as
\begin{equation}\label{eq:definitionfluxes}
	\begin{split}
		M_i=\frac{1}{(2\pi)^2 \alpha'}\int_{\mathcal{A}_i} F_3 \quad\text{ and }\quad K_i=-\frac{1}{(2\pi)^2 \alpha'}\int_{\mathcal{B}_i}H_3\fstop
	\end{split}
\end{equation}
As shown in~\cite{Giddings:2001yu,Sethi:1996es}, generic choices of the fluxes stabilize the c.s. moduli and the axiodilaton. Furthermore, if $K_i\gg g_s M_i$, the c.s. modulus associated to the cycle $\mathcal{A}_i$ will be stabilized close to the conifold point,
\begin{equation}
\left|z_i\right|=\exp\left(-2\pi \dfrac{K_i}{g_s M_i}\right)\ll 1\fstop
\end{equation}
\noindent Everywhere in this paper we will work under this assumption, which we request to hold independently for each pair of flux quanta $K_i$, $M_i$.

When the c.s. moduli are stabilized close to the conifold point, the multi-conifold system described in the previous paragraphs is replaced with a system of long multiple throats, which arise due to the backreaction of fluxes on the geometry.
Within the throats, the metric is well approximated by the Klebanov-Tseytlin~\cite{Klebanov:2000nc} solution
\begin{equation}
ds^2=w(r)^2\eta_{\mu\nu}dx^{\mu}dx^{\nu}+w(r)^{-2}\left(dr^2+2^2ds_{T_{1,1}^2}\right)\coma w(r)^2\sim \dfrac{r^2}{g_s M\alpha'}\log\left(\frac{r}{r_{\IR}}\right)^{-\frac{1}{2}}\coma
\end{equation}
where $r$ is the radial coordinate, $w(r)$ is the warp factor, and $M$ is the flux quanta defined in~\eqref{eq:definitionfluxes}. The solution ceases to hold at a UV cutoff $r_{\UV}$, where the multi-throat is attached to the bulk geometry, and also at a IR cutoff $r_{\IR}$ near the bottom of the throats. For $r\lesssim r_{\IR}$ the metric is given by the full Klebanov-Strassler (KS) solution~\cite{Klebanov:2000hb}. An exponential hierarchy, as the one in Randall-Sundrum model~\cite{Randall:1999ee,Giddings:2001yu}, is thus engineered by $w_{\IR}\equiv w(r_{\IR})\sim r_{\IR}/r_{\UV}\sim |z^{1/3}|$.

We will now review in more detail how the c.s. moduli stabilization operates, in this setup. For concreteness, we focus on a subset of $n\leq h^{2,1}_-$ c.s. moduli associated to $n$ $\mathcal{A}$-cycles subject to $m$ homology relations. The superpotential coupling the thraxion fields $G^I$ to the c.s. moduli can be derived from the Gukov-Vafa-Witten (GVW) superpotential~\cite{Gukov:1999ya} and it reads~\cite{Hebecker:2018yxs}
\begin{equation}\label{eq:superpotential}
	W=\sum_{k=1}^{n}\left(M_k \frac{z_k}{2\pi i}\log z_k-\tau K_k z_k\right)-\sum_{I=1}^{m}\frac{G^I}{2\pi} \mathcal{P}_I+\hat{\hat{W_0}}(z) +\mathcal{O}(z_k^2)\coma
\end{equation}
where $\hat{W_0}(z)$ is a holomorphic function denoting contributions from other cycles and $\mathcal{P}_I$ are the $m$ relations for the $n$ c.s. moduli $z_i$, $\mathcal{P}_I  \equiv \sum_{k=1}^n \sum_{I=1}^{m} p_I^k z_k\,$.

In \eqref{eq:superpotential} the $n$ complex structure moduli are thought to be all independent. The fact that they are subject to $m$ relations is imposed dynamically, once the thraxions $G^I$ are set on-shell by their equation of motion. In particular, the fields $G^I$ act as Lagrange multipliers, imposing the homology relations among the c.s. moduli.

On the other hand, the K\"ahler potential for the complex structure moduli reads
\begin{equation}
\begin{aligned}
K_{\text{c.s.}}(z_i, \bar{z}_i)&=-\log\left(-i\int \Omega\wedge \bar{\Omega}\right)=-\log\left(ig_K(z)-\overline{g_K(z)}+\sum_{I=1}^{n-m}i\bar{z}_IG^I+\text{ c.c.}\right)=\\
&=-\log\left(ig_K(z)-\overline{g_K(z)}+\sum_{i=1}^{n}\left[\dfrac{|z_i|^2}{2\pi}\log(|z_i|^2)+i\bar{z}_ig^i(z)-iz_i\overline{g^i(z)}\right]\right)\coma
\end{aligned}
\label{eq:Kcsthr}
\end{equation}
where $g_K(z)$ is a holomorphic function encoding contributions from the periods of $h^{2,1}_--n$ c.s. moduli of the bulk CY, while the $g_i(z)$ are related to the periods of the c.s. moduli of the multi-throat. Being interested in small $z_i$ we can Taylor expand these functions, giving
\begin{equation}
g_k(z)=g_{K,0}+\sum_{j}g_{K,1}^{i}z_i+\mathcal{O}(z^2)\coma \,
g^i(z)=g_0^i+\sum_{j}g^{ij}z_j+\mathcal{O}(z^2)\fstop
\end{equation}
It is also possible to expand $\hat{\hat{W}}(z)$ as
\begin{equation}
    \hat{\hat{W}}(z)=g_{W,0}+\sum_{i=1}^ng^1_{W,1}z_i+\mathcal{O}(z^2)\coma
\end{equation}
and define 
\begin{equation}
    \hat{W}_0\equiv g_{W,0}+\sum_{i=1}^nM_ig_0^i
\end{equation}
to be the superpotential containing all the contributions of order $\mathcal{O}(z^0)$. 
By computing the F-term equations of the c.s. moduli $z_i$ using \eqref{eq:superpotential}, one can relate the VEV of the $n$ c.s. moduli to the VEV of the thraxions, as\footnote{In Eq.\eqref{eq:valuesofzi} we have already set $C_0$ to zero.}
\begin{equation}\label{eq:valuesofzi}
\begin{split}  
    &z_{k}=z_{0,k}\,e^{ i \sum_I \frac{p_k^IG^I}{M_k}}\quad \text{where }\\ 
    &\,\,z_{0,k}=e^{-1-2\pi\frac{K_k}{g_s M_k}}e^{-\frac{2\pi i}{M_k}\left(\sum_j g^{kj}_1+g^k_{W,1}-i\frac{\tilde g_0^k \hat{W}_0}{2 \im{g_{K,0}}}\right)}+\mathcal{O}\left(e^{-4\pi\frac{K_k}{g_sM_k}}\right)\fstop
\end{split}
\end{equation}
We remark that at the current level of the discussion, the thraxion fields themselves are \emph{not} stabilized yet. Therefore, the c.s. moduli themselves are yet not stabilized, but simply expressed in terms of the VEV of the thraxion and flux quanta. By plugging \eqref{eq:valuesofzi} in \eqref{eq:superpotential}, we find the effective superpotential for the thraxions we will be using in the stabilization procedure.
Let us consider the effective thraxion superpotential for $n$ throats and $m$ thraxions~\cite{Hebecker:2018yxs}:
\begin{equation}
    W_{\text{eff}}=\hat{W}_0-\sum_{k=1}^n\epsilon_{k}e^{i\sum_{I=1}^mp_I^kG^I/M_k}\coma
    \label{eq:thraxionsuperpot}
\end{equation}
where
\begin{equation}
    \epsilon_k\equiv \frac{M_k}{2\pi i}z_{0,k}\left(1-\frac{2\pi}{M_k}\frac{\hat{W}_0\overline{\tilde{g}_0^k}}{\mathfrak{a}}\right)\fstop
    \label{eq:genepsilon}
\end{equation}
We also introduced
\begin{equation}
    \tilde{g}_0^k=g_0^k-\overline{g^k_{K,1}}\coma \mathfrak{a}\equiv -2\im(g_{K,0})
\end{equation}
and $z_{0,k}$ given in Eq.~\eqref{eq:valuesofzi}. We remark that the physical deformation parameters are $z_{k}$ defined in Eq.~\eqref{eq:valuesofzi}. As we will analyze in Section~\ref{sec:ThraxionsTorclass}, a VEV different from zero for $G^I$ is necessary to generate a potential for the thraxions, but whenever they are not stabilized at zero, the CY condition is broken. Indeed, generically $G^I$ does not need to stabilize at zero. Whenever the thraxions do not stabilize at vanishing VEV, they induce backreaction that breaks the CY condition for the extra dimensions. We will comment about this at length in Section~\ref{sec:ThraxionsTorclass}.%

By using some approximations, one can simplify Eq.~\eqref{eq:thraxionsuperpot} above. In particular, in the explicit examples discussed in Section~\ref{sec:ModuliStabilization} and in Appendix~\ref{app:modstabdetails} we will always work with a simplified superpotential. As used in~\cite{Hebecker:2018yxs}, supposing that only $g_{W,0}$ and $g_{K,0}$ are non-vanishing, the definitions of $z_{0,k}$, $\epsilon_k$ and $\hat{W}_0$ simplify.\footnote{Such approximation imposes that $\hat{W}_0$ and all the non-logarithmic terms in Eq.~\eqref{eq:Kcsthr} are constants.} For the case in which there is only one thraxion in one multi-throat system composed of $n$ throats, by using a symmetrical choice of $M_k$ and $K_k$ fluxes, it is possible to rewrite Eq.~\eqref{eq:thraxionsuperpot} as
\begin{equation}
\label{eq:1dtsuperpot}
 	W_{\text{thr}}(G)=W_0+n_p \epsilon \left(1-\cos\left(G/M\right)\right)\coma
\end{equation}
where $n_p$ is the number of KS throats in the multi-throat system, and we have defined $W_0\equiv \hat{W}_0-n_p\,\epsilon$.\footnote{In this way, the axion $c$ has the well-known axion effective potential.} In the case in which there are $k$ multi-throats in the CY, each one hosting a single thraxion the superpotential can approximately be written as $q$ copies of \eqref{eq:thraxionsuperpot}, namely 
\begin{equation}
\label{eq:dtsuperpot}
 	W_{\text{thr}}(G^I)=W_0+\sum_{I=1}^q \epsilon_I \left(1-\cos\left(G^I/M_I\right)\right)\coma
\end{equation}
 where we have absorbed the number of KS throats inside each system in the definition of $\epsilon_I$. 

So far we reviewed the problem of moduli stabilization for the c.s. moduli, in presence of thraxions. One is then left with discussing the problem of moduli stabilization for the thraxions themselves and for the K\"ahler moduli. The total K\"ahler potential actually reads
\begin{equation}
\label{eq:Kahlertot}
	K\left(G,\bar{G},T,\bar{T},z,\bar{z}\right)=K_{\text{c.s.}}(z,\bar{z})+K_{\text{thr}}(G-\bar{G},T+\bar{T})\coma
\end{equation}
where $K_{\text{c.s.}}$ has been introduced in~\eqref{eq:Kcsthr}, while $K_{\text{thr}}$ is a K\"ahler potential coupling the thraxions to K\"ahler moduli. In \eqref{eq:Kahlertot} we introduced the complexified K\"ahler moduli\footnote{We have introduced the index $a$ for the fields $G^a$ because in general in the resolved side of the conifold transitions not all the $h^{1,1}_-$ moduli come from the presence of thraxions. We can call $\tilde{h}^{1,1}_-$ the moduli that do not come from thraxions, such that $h^{1,1}_- = m + \tilde{h}^{1,1}_-$, hence in principle we should have $a=1,\ldots,m+\tilde{h}^{1,1}_-$. For the purpose of this paper, we will always assume $\tilde{h}^{1,1}_-=0$.
}
\begin{equation}
    \begin{split}
    T_\alpha&= \tau_\alpha + i \theta_\alpha +\frac{i\,\kappa_{\alpha ab}}{2\left(\tau-\bar{\tau}\right)}G^a\left(G-\bar{G}\right)^b \\
    &=\left(\tau_\alpha-\frac{1}{2g_s}\kappa_{\alpha ab}b^ab^b\right)+i\left(\theta_\alpha
    -\frac{1}{2}\kappa_{\alpha ab}c^ab^b\right)\coma
    \end{split}
    \label{eq:Tfieldsdef}
\end{equation}
where $\kappa_{\alpha ab}$ are the triple intersection numbers between the $\alpha$-th orientifold-even 4-cycle and the orientifold-odd combination of a pair of orientifold swapped $4$-cycles, 
$\alpha=1,\dots,h_{+}^{1,1}$ and 
\begin{equation}
    \tau_\alpha + i \theta_\alpha\coloneqq\frac{1}{2}\kappa_{\alpha\beta\gamma}t^\alpha t^\beta+i \int_{D^\alpha}C_4\fstop
\end{equation}
Notice that in the second line of~\eqref{eq:Tfieldsdef} we have already set $C_0$ to zero. We assume that one can invert the relation between the 2-cycle moduli and the 4-cycle moduli to write the K\"ahler potential $K_{\text{thr}}$ for the $T$ and $G$ fields as 
\begin{equation}\label{eq:kpotthraxions}
 	\begin{split}
 		&K_{\text{thr}}=-2 \log\left(F\right)\coma \mbox{where}\\
 		&F=\sum_{\alpha=1}^{h_+^{1,1}}\sum_{a,b=1}^{h_-^{1,1}}c_\alpha\left(T_\alpha+\bar{T}_\alpha-\frac{g_s}{4}\kappa_{\alpha ab}\left(G^a-\bar{G}^a\right)\left(G^b-\bar{G}^b\right)\right)^{3/2} \fstop
 	\end{split}
 \end{equation}
 However, we stress that the discussion we will carry out in Sections~\ref{sec:structurePotgen} and~\ref{sec:structurePotdt} does not need to assume any explicit expression for the K\"ahler potential. 
 
The F-term $4$d supergravity potential can be computed as
\begin{equation}
    V=e^{K}\left[K^{A\bar{B}}\mathcal{D}_A W\mathcal{D}_{\bar{B}}\bar{W}-3|W|^2\right]\coma
    \label{eq:VSUGRA}
\end{equation}
where $\mathcal{D}_AW\equiv\partial_A W + K_A W$ is the K\"ahler covariant derivative and the indices $A,B$ run over all the moduli $T_\alpha$ and $G^a$.
 Notice that the no-scale relation for $K_{\text{thr}}$, $$\partial K_{\text{thr}}^{\dagger}\left(\partial^2 K_{\text{thr}}\right)^{-1}\partial K_{\text{thr}}=3\coma$$ is satisfied. Let us consider for simplicity the case of a single thraxion. Thanks to the no-scale property of $K_{\text{thr}}$, one can show that the F-term potential scales as
\begin{equation}\label{eq:ISDthraxpot}
    V\left(G,\bar{G}\right)\propto |\partial_G W\left(G\right)|^2 \fstop
\end{equation}
Hence, the potential for the thraxion gets a double suppression in the $\epsilon\sim z_0\sim \omega_{\IR}^3$ parameter. In turn, by construction this implies that the mass-squared of the $G$ field is of order $\omega_{\IR}^6\ll 1$, making the thraxion an extremely light particle. This effect generalizes trivially for the case of multiple thraxions.

So far, the K\"ahler moduli sector is left as a flat direction of the potential. We have considered only tree-level contributions to the superpotential, which come from the presence of thraxions in the theory. In Section~\ref{sec:ModuliStabilization} we will study if and how the potential for thraxions gets modified by the inclusion of perturbative and non-perturbative quantum effects proportional to the K\"ahler moduli.

\section{Backreaction of Thraxions on $\SU(3)$ Torsion Classes}
\label{sec:ThraxionsTorclass}

As it is well known, the presence of fluxes and localized objects backreacts on spacetime, generically causing the compactification manifold $\IX$ to cease to be CY, yet still maintaining a $\SU(3)$ structure. Geometrical properties of the backreacted compactification manifold can be understood by an analysis of its $\SU(3)$ torsion classes. In this section, we focus on the effect that the presence of non-vanishing thraxion VEVs produces on the torsion classes of $\IX$, in a IIB compactification. We follow the exposition and conventions of~\cite{Grana:2005jc} and we refer to that review (and references therein) for details.

Let $\IX$ be a $\SU(3)$ structure manifold. Notice that this in particular implies that, since $\SU(3)\subset \SO(6)$, $\IX$ is a particular case of a $\SO(6)$ structure manifold. Suppose that the metric-compatible connection on $\IX$ is the Levi-Civita connection $\nabla'_m$, which is torsionless. Then $\IX$ is CY
if and only if it admits a non-vanishing globally well-defined spinor which is also covariantly constant. However, in general, it is possible that the metric-compatible connection for the $\SU(3)$ structure manifold is not the Levi-Civita one, and in particular it might have torsion. 

Let $\nabla'_m$ be a generic metric-compatible connection. Its Lie bracket, acting on a vector $V_p\in \Gamma(TM)$ will read
\begin{equation}
    \left[\nabla'_m,\nabla'_n\right]V_p=-R_{mnp}^qV_q-2T_{mn}^q\nabla'_qV_p\coma
\end{equation}
where $R_{mnp}^q$ is the Riemann curvature tensor, and $T_{mn}^q$ is the torsion tensor. We take this as the definition of the torsion tensor. We notice that $T^p_{mn} \in \Gamma(TM\otimes \Omega^2(M))$, namely, the torsion tensor is a section of the space of 1-vectors tensored with the space of $2$-forms. Viewing $\Omega^2(M)$ as a vector space, it holds $\Omega^2(M)\simeq \mathfrak{so}(6)$, the 15-dimensional vector space on which the Lie algebra $\mathfrak{so}(6)$ of the structure group $\SO(6)$ acts. Similarly, $TM$ can be thought as a 6-dimensional vector space on which the fundamental representation of $\mathfrak{so}(6)$ acts.

Since $\IX$ has not only $\SO(6)$ structure, but more specifically $\SU(3)\subset \SO(6)$ structure, it is possible to split the vector spaces $TM$ and $\Omega^2(M)$ in subspaces that are left invariant by the action of $\mathfrak{su}(3)$. This amounts to find the branching rules of the adjoint $15$-dimensional representation of $\mathfrak{so}(6)$, acting on $\Omega^2(M)$, and the fundamental $6$-dimensional representation of $\mathfrak{so}(6)$, acting on $TM$. This is
\begin{equation}
\label{branching}
\begin{aligned}
\mathbf{15}&\to \mathbf{8}\oplus\mathbf{3}\oplus\mathbf{\bar{3}}\oplus \mathbf{1}\\
\mathbf{6}&\to\mathbf{3}\oplus\mathbf{\bar{3}}
\end{aligned}
\end{equation}
From \eqref{branching} we see that $\Omega^2(M)\simeq \mathfrak{so}(6)$ splits in the direct sum of an 8-dimensional vector space on which the $8$-dimensional adjoint representation of $\mathfrak{su}(3)$ acts, plus a $7$-dimensional vector space on which $\mathbf{3}\oplus\mathbf{\bar{3}}\oplus \mathbf{1}$ acts. 

In conclusion, we find that $T_{mn}^p$ is a section of $(\mathbf{3}\oplus\mathbf{\bar{3}})\otimes(\mathbf{8}\oplus\mathbf{3}\oplus\mathbf{\bar{3}}\oplus \mathbf{1})$, where now we are committing the small abuse of notation of denoting the vector space on which a given representation of a Lie algebra acts, with the representation itself.

In order to define the torsion classes, we need to focus on the components of the torsion tensor which, under a generic $\SU(3)$ action, leave invariant the $\mathbf{8}$ subspace, i.e.
\begin{equation}
     T^0_{mn}{}^p\in (\bf{3}\oplus\overline{\bf{3}})\otimes (\bf{3}\oplus \overline{\bf{3}}\oplus \bf{1})\fstop
     \label{eq:intrinsictorsiondef}
\end{equation}
One then defines the torsion classes as a particular reducible representation (or equivalently the vector spaces on which they act) in the expression \eqref{eq:intrinsictorsiondef}, namely

\begin{equation}
\begin{array}{ccccccccc}
    T^0_{mn}{}^p & = & (\bf{3}\oplus\overline{\bf{3}})&\otimes &(\bf{3}\oplus \overline{\bf{3}}\oplus \bf{1}) \\
    & = & (\bf{1}\oplus\bf{1}) &\otimes & (\bf{8}\oplus\bf{8}) & \otimes & (\bf{6}\oplus\overline{\bf{6}}) &\otimes & 2(\bf{3}\oplus\overline{\bf{3}})\\\
     &  & W_1 & & W_2 & & W_3 & & W_4,\, W_5
     \label{eq:intrinsictorsion}
\end{array}
\end{equation}
We have introduced in \eqref{eq:intrinsictorsion} the torsion classes $W_i$, $i=1,\ldots,5$.

Following~\cite{Chiossi:2002aa,Grana:2004bg,Grana:2005jc}, they transform in the following way:
\begin{itemize}
    \item $W_1$ is a complex scalar.
    \item $W_2$ is a complex primitive $(1,1)$-form.
    \item $W_3$ is a real primitive $(2,1)+(1,2)$-form.
    \item $W_4$ is a real vector.
    \item $W_5$ is a complex $(1,0)$-form.
\end{itemize}
Here, we recall that a differential form $\omega$ is primitive if $\omega \wedge J=0$, where $J$ is the K\"ahler form of $\IX$. Depending on which torsion classes vanish (or take specific values), different properties of $\IX$ can be present or absent, such as for example the symplectic, K\"ahler, or CY structure. $\SU(3)$ structure manifolds can be organized in this way in nine different classes, which we recall in Table~\ref{tab:torsionclasses}.

\begin{table}[!htp]
    \centering
    \renewcommand{\arraystretch}{1.25}
    \begin{tabular}{c|c}
        \textbf{Manifold} & \textbf{Vanishing torsion classes} \\
        \hline
        Complex & $W_1=W_2=0$\\\hline
        Symplectic & $W_1=W_3=W_4=0$\\\hline
        Half-flat & $\im W_1=\im W_2=W_4=W_5=0$\\\hline
        Special Hermitian & $W_1=W_2=W_4=W_5=0$\\\hline
        Nearly K\"ahler & $W_2=W_3=W_4=W_5=0$\\\hline
        Almost K\"ahler & $W_1=W_3=W_4=W_5=0$\\\hline
        K\"ahler & $W_1=W_2=W_3=W_4=0$\\\hline
        Calabi-Yau & $W_1=W_2=W_3=W_4=W_5=0$\\\hline
        Conformal Calabi-Yau & $W_1=W_2=W_3=3W_4-2W_5=0$
    \end{tabular}
    \caption{Vanishing torsion classes in special $\SU(3)$ structure manifolds~\cite{Grana:2005jc}.}
  \label{tab:torsionclasses}
\end{table}

We now focus on $4$d $\mathcal{N}=1$ compactifications of type IIB superstring, where we take the external space to be $\mathbb{R}^{1,3}$. We make the ansatz that the $10$d metric is
\begin{equation}
    ds^2=e^{2A(y)}\eta_{\mu\nu}dx^\mu dx^\nu+g_{mn}dy^mdy^n \coma \text{with } \mu,\nu=0,\ldots,3\coma m,n=1,\ldots,6\coma 
    \label{eq:10dmetric}
\end{equation}
where $\eta_{\mu\nu}$ is the $4$d Minkowski metric, and $g_{mn}$ is the metric of the internal manifold $\IX$. We have also introduced a warp factor $A(y)$, a function of only the internal coordinates.

We first split the ten-dimensional supersymmetry spinors into $4$-dimensional and $6$-dimensional ones~\cite{Grana:2005jc}:
\begin{equation}
    \epsilon^1=\xi_+^1\otimes \eta_++\xi^1_-\otimes \eta_- \coma \epsilon^2=\xi_+^2\otimes \eta_++\xi^2_-\otimes \eta_-\coma
    \label{eq:IIBN=2spinors}
\end{equation}
where $\eta_{\pm}$ are the covariantly constant spinors of the $\IX$ manifold, i.e.
\begin{equation}
    \nabla_m \eta_\pm =0\fstop
\end{equation}
We introduce two complex functions of the internal coordinates $a(y)$ and $b(y)$, such that $|a|^2+|b|^2=e^A$. The existence of such functions is a consequence of the request that the vacuum is $\mathcal{N}=1$. These functions parametrize the way in which a given $\mathcal{N}=1$ subalgebra is embedded in the larger $\mathcal{N}=2$ SUSY algebra. The spinors in Eq.~\eqref{eq:IIBN=2spinors} then become
\begin{equation}
    \epsilon^1=\xi_+^1\otimes (a\eta_+)+\xi^1_-\otimes (\overline{a}\eta_-) \coma \epsilon^2=\xi_+^2\otimes (b\eta_+)+\xi^2_-\otimes (\overline{b}\eta_-)\fstop
\end{equation}
It is possible to obtain general expressions for the torsions, the fluxes, the warp factor of the metric and the functions $a$ and $b$ that classify all possible $\mathcal{N}=1$ vacua in type IIB. From~\cite{Grana:2004bg,Grana:2005jc} we take the following relations which are valid for any $4$d Minkowski vacuum in type IIB:
\begin{equation}
    \begin{split}
        2 ab W_3 &= e^{\phi}  (a^2+b^2)   F^{(6)}_{3}\coma\\
        (a^2 - b^2) W_3 &= - (a^2+b^2) \star_6H_3^{(6)}\coma\\
        2 ab H_3^{(6)} &= - e^{\phi} (a^2 - b^2)\star_6 F^{(6)}_{3}\coma 
    \end{split}
    \label{eq:6int}
\end{equation}
\begin{equation}
  \label{eq:intvec}
  \renewcommand{\arraystretch}{2}
    \begin{array}{lll}
e^{\phi} F^{(\bar 3)}_3 &=& 
\dfrac{-4i  ab (a^2+b^2) }{a^4-2ia^3b+2iab^3+b^4}   \bar \de  a
 \, , \\
e^{\phi} F^{(\bar 3)}_5 &=& 
\dfrac{-4  ab (a^2-b^2) }{a^4-2ia^3b+2iab^3+b^4}   \bar \de  a\, , \\
H_3^{(\bar 3)} &=& 
 \dfrac{-2i  (a^2 + b^2) (a^2-b^2)}{a^4-2ia^3b+2iab^3+b^4}   \bar \de a  ,\\
 \bar \partial  A &=& - \dfrac{4 (ab)^2}{a^4-2ia^3b+2iab^3+b^4}\bar \de a \, ,  \\
\bar \de \phi &=&   \dfrac{2 (a^2 + b^2)^2 }{a^4-2ia^3b+2iab^3+b^4} \bar \de a  \, ,
\end{array}\quad
\renewcommand{\arraystretch}{2}
\begin{array}{lll}
W_1 &=& 0\coma \\
W_2 &=& 0\coma \\
W_4 &=&  \dfrac{ 2 (a^2 - b^2)^2}{a^4-2ia^3b+2iab^3+b^4 } \bar \de a \, ,  \\
\bar W_5 &=&  \dfrac{2( a^4-4 a^2 b^2 +b^4)}{a^4-2ia^3b+2iab^3+b^4} \bar \de a \, ,
\end{array}
\end{equation}
where $\phi$ is the $10$d dilaton, and with the superscript we indicate the dimension of the representation of $\SU(3)$. GKP-like solutions~\cite{Giddings:2001yu} correspond to $a=\pm ib$. In this case \eqref{eq:6int} and \eqref{eq:intvec} will read:
\begin{equation}
    \begin{split}
    W_1=0\,, &\quad W_2=0\coma W_3=0 \coma \bar{\de}\phi=0 \coma e^\phi F_3^{(6)}=\mp \star_6 H_3^{(6)}\coma\\
    &e^\phi F_5^{\left(\overline{3}\right)} =\frac{2}{3}i\overline{W}_5=iW_4=-2i\bar{\de}A=-4i\bar{\de}\log a\fstop
    \end{split}
    \label{eq:GKPtorsolu}
\end{equation}
The condition $e^\phi F_3^{(6)}=\mp \star_6 H_3^{(6)}$ is equivalent to the request that
$ G_3=F_3-ie^{-\phi}H_3$
is ISD and has no single $(0,3)$ component:
\begin{equation}
    \star_6 G_3=iG_3 \coma G_{(0,3)}=0\fstop
\end{equation}
The last condition in \eqref{eq:GKPtorsolu} is implying that the manifold is conformal CY, since all torsion classes are zero except for $2W_5=3W_4$.

From Eq.~\eqref{eq:GKPtorsolu}, we see that on a conformal CY $W_4$ and $W_5$ are only sourced by 5-form flux $F_{5}^{(3)}$. On the other hand from Eq.~\eqref{eq:6int}, we have that only the $(2,1)$-form fluxes $H_3^{(6)},F_3^{(6)}$ can source $W_3$ and that depending on the choice of the preserved $4$d spinor pair, either the ISD combination of $H_3^{(6)},F_3^{(6)}$ (corresponding to the choice $a=ib$) or the anti-ISD ($a=-ib$) will set $W_3=0$.

It does now become clear what the drastic effect of a non-zero thraxion VEV on a warped (conformal) CY is. For this, we recall that in order for a thraxion to exist there must be present in $\IX$ at least one warped multi-throat region, with at least one homology relation among the shrinking $\beta$-cycles.

This setting implies the presence of quantized $(2,1)$-form background fluxes. Hence, there is an amount of $ \left.H_3^{(6)}\right|_0, \left.F_3^{(6)}\right|_0$ stabilizing the whole c.s. moduli sector and warp factor.
On top of this background flux configuration, turning on the thraxion corresponds to turning on `a bit' of $C_2$ at the IR ends of the multi-throat, with profile in the throat radial direction. Thus, turning on the thraxion corresponds to turning on `a bit' of pure\footnote{The thraxion-induced $F_3$-flux lives on a $3$-cycle in the throat part of the $(2,1)$-homology of the CY, and thus has to be locally of the same cohomology type as the ISD background fluxes.} $\Delta F_3^{(6)}$ but \emph{without} any accompanying $H_3^{(6)}$. 

Hence, the extra thraxion-flux $\Delta F_3^{(6)}$ is non-ISD. The reason for which the extra thraxion-flux breaks the ISD condition is that the c.s. moduli of the multi-throat simply cannot adjust their VEVs in order to be ISD again with respect to the new configuration. This is impossible for the following reason. If they could adjust their VEVs to be ISD again, this would imply that the thraxion potential vanishes. However, this was shown to be impossible since the $10$d equations of motion of the perturbed multi-throat analyzed in~\cite{Hebecker:2018yxs} forbid it, once the multi-throat is embedded in a compact CY. The system cannot relax back to vanishing vacuum energy at finite thraxion-flux. Schematically, this can be denoted as:
\begin{equation}
\left(\text{ISD}\;\; \Rightarrow\;\; V=0\land \partial_{z^i}V=0\;\forall i\right)\quad\Rightarrow\quad\left(V\neq0\lor  \partial_{z^i}V\neq 0\;\;\Rightarrow \;\;\text{ non-ISD}\right)\,.
\end{equation}

We now analyze the effect of this violation of ISD on the torsion classes of the compactification. For this purpose, we keep treating the solution as it was living in $4$d Minkowski space, neglecting the tiny positive effective Cosmological Constant (CC) of the thraxion potential. We argue that this approximation makes sense as the size of this potential is parametrically small due to warp factor suppression. Furthermore, the CC-induced gravitational backreaction is also suppressed by the $1/M_{\rm  P}^2$ gravitational coupling of the Einstein equations. 

As discussed above, the main backreaction from the thraxion will be its non-ISD nature, which distorts the torsion classes of the manifold. In glossing over the multiple $3$-cycles of an actual CY we can see, that by writing this extra thraxion-flux $\Delta F_3^{(6)}\equiv\epsilon  \left.F_3^{(6)}\right|_0$ the ISD relation \eqref{eq:6int} changes as
\begin{equation}
\begin{split}
    2ab H_3^{(6)}&=2 a (b_0+\delta b)\left.H_3^{(6)}\right|_0=-e^\phi (a^2-(b_0+\delta b)^2)\star_6\left.F_3^{(6)}\right|_0\\
    &=-e^\phi (a^2-b_0^2)\star_6\left.F_3^{(6)}\right|_0+e^\phi \epsilon \star_6\left.F_3^{(6)}\right|_0\,.
\end{split}
\end{equation}
Plugging in $a=ib_0$ this becomes
\begin{equation}
\begin{split}
    2ib_0^2\left.H_3^{(6)}\right|_0+2ib_0\delta b \left.H_3^{(6)}\right|_0&=e^\phi(2b_0^2+2b_0\delta b+\delta b^2)\star_6\left.F_3^{(6)}\right|_0\\
    &=e^\phi 2b_0^2 \star_6\left.F_3^{(6)}\right|_0+e^\phi\epsilon \star_6\left.F_3^{(6)}\right|_0\,.
\end{split}
\end{equation}
Canceling out identical pieces, we see that $\delta b=\sqrt\epsilon$, that is, the thraxion-sourced extra $\Delta F_3^{(6)}={\cal O}(\epsilon)$ deforms the ISD relation $a=ib_0$ to $a\neq ib$ with deformation ${\cal O}(\epsilon^{1/2})$. Hence, the flux with turned-on thraxion is non-ISD in such a way, that $W_3\neq0$ because now $a\neq \pm ib$. According to Table \ref{tab:torsionclasses} the thraxion thus `wrecks' the CY in the qualitatively worst fashion -- leaving just a complex manifold.

Still, the extreme scale suppression of the thraxion sector due to warping may leave this just-complex non-CY manifold in some sense ``near" the original conformal CY, where the word ``near" awaits an appropriate definition of distance in torsion deformation space and the space of 4d effective actions from KK reduction, which is beyond the scope of this paper.

\section{Moduli Stabilization}
\label{sec:ModuliStabilization}
The GKP-type flux compactification that we considered so far discusses the stabilization of c.s. moduli in presence of thraxions. We are still left with the problem of K\"ahler moduli stabilization, which we will address in this section. Whenever thraxions are present, there are two sources of no-scale breaking, which contribute to K\"ahler moduli stabilization: one is the usual F-term scalar potential coming from the introduction of non-perturbative corrections to the superpotential, while the other one is the CY breaking potential of the thraxions reviewed in Section~\ref{sec:reviewThraxions}. In \cite{Hebecker:2018yxs}, an initial study of the mixing between these two effects was discussed. However, no detailed analysis was carried out. 

 The aim of this section is to perform this combined analysis. In particular, we will study the backreaction on the thraxion potential~\eqref{eq:ISDthraxpot} when we appropriately stabilize the K\"ahler moduli via the leading stabilization mechanisms. In particular, we will answer whether K\"ahler moduli stabilization can change the property that the thraxion mass squared is six-time suppressed by the warping.
 As a main result, we find a 2-fold statement.  On one hand, \emph{generically} the six-fold warp suppression is spoiled, once the thraxion-carrying multi-throat consists of at least 3 connected throats. On the other hand, for double-throats there exist classes of CY flux compactifications where the six-times warp suppression survives K\"ahler moduli stabilization.
 
The survival of the full warp suppression depends on the K\"ahler moduli stabilization, potentially inducing a cross term which is proportional to the warp factor cubed only. Consequently, the mass gets lifted whenever this cross term does not vanish. The mass squared of the warped-throat KK modes scales as $\omega_{\IR}^2$. Moreover, we will show below that while the stabilized K\"ahler moduli are parametrically lighter than the warped KK modes, the thraxion is still parametrically lighter than the K\"ahler moduli. Hence, it remains the lightest state inside the throat. The mass-spectrum is still \emph{effectively} gapped.
 
 We show how this works in the most general case in Section~\ref{sec:structurePotgen}. Nevertheless, we find that in certain classes of setups the $\mathcal{O}(\epsilon)$ cross term (which is responsible for lifting the mass) gets canceled by K\"ahler moduli stabilization. This happens when we set the $C_4$ axion to its minimum.\footnote{We will also need some assumption on the VEV of the saxion $b$. We will comment on this point later on.} In Section~\ref{sec:structurePotdt} we explain in which cases this holds and which is the amount of tuning required. Moreover, in Sections~\ref{sec:structurePotdt} and~\ref{sec:MS-LVS} we specialize the discussion to KKLT and LVS respectively.

The flatness of the F-term scalar potential for the K\"ahler moduli can be cured by considering perturbative and non-perturbative corrections to the K\"ahler potential and the superpotential. For the superpotential, only non-perturbative ones are allowed in the K\"ahler moduli $T$. These corrections can be generated either by Euclidean D$3$-brane (ED$3$-brane) instantons~\cite{Uranga:2008nh} or by gaugino condensation effect happening in the worldvolume theories of stacks of D$7$-branes wrapping rigid divisors~\cite{Gorlich:2004qm,Witten:1996bn}. Both these contributions take the form
\begin{equation}\label{eq:npsuperpot}
    W_{\text{np}}= \sum_{\alpha} A_{\alpha}e^{-a_\alpha T_\alpha}\coma
\end{equation}
where $a_\alpha =2\pi$ for ED$3$-branes and $a_\alpha=\frac{2\pi}{h^{\vee}(G_\alpha)}$ for the gaugino condensation case. Here $h^{\vee}(G_\alpha)$ is the dual Coxeter number of the gauge group $G_\alpha$ on the $\alpha$-th stack of D$7$-branes. The coefficients $A_\alpha$ depend on the stabilization of the c.s. moduli. Additionally, there might be higher instanton corrections, but these can be neglected as long as $a_\alpha T_\alpha>1$. 

The K\"ahler potential admits also perturbative corrections both in $g_s$ and $\alpha'$. In the following, we neglect string-loop corrections. 
 The leading $\alpha'$ correction of the 4d  $\mathcal{N}=2$ effective action reads~\cite{Becker:2002nn} 
\begin{equation}
K=-2 \ln\left(\mathcal{V}+\frac{\hat{\xi}}{2 }\right)\coma
\label{eq:KpotBBHL}
\end{equation}
where $\hat\xi$ is a constant which controls the strength of $\alpha'$ corrections and is given by $\hat\xi=\xi/g_s^{3/2}$, $\xi=-\frac{\chi\left(\IX\right)\zeta\left(3\right)}{ 2\left(2\pi\right)^3}$, $\zeta$ being the Riemann zeta function and $\chi\!\left(\IX\right)=2\left(h_{1,1}-h_{2,1}\right)$ being the Euler number of the CY 3-fold $\IX$.
In the $\mathcal{N}=1$ case, this correction was shown to be the only one to contribute thanks to the extended no-scale cancellation~\cite{Cicoli:2007xp} of $\mathcal{O}(\alpha'^2)$ and the subdominance of higher terms.\footnote{Notice that in the $\mathcal{N}=1$ case with O$7$-planes one should trade $\chi(\IX)$ with $\chi_{\text{eff}}(\IX)=\chi(\IX)+2\int_{\IX} D_{O7}^3$ in the definition of $\hat\xi$~\cite{Minasian:2015bxa}.}

Let us specialize the F-term potential in Eq.~\eqref{eq:VSUGRA} by including the perturbative and non-perturbative corrections of Eqs.~\eqref{eq:KpotBBHL} and~\eqref{eq:npsuperpot}. Hence, we can rewrite it
as~\cite{Balasubramanian:2004uy,Balasubramanian:2005zx}
\begin{equation}
\begin{split}
    V = &\, e^K\left[\sum_{\alpha,\beta} K^{T_\alpha\bar{T}_\beta}a_\alpha A_\alpha a_\beta A_\beta e^{-a_\alpha T_\alpha-a_\beta\bar{T}_\beta}\right]+\\
    & -e^K\left[\sum_{\alpha,\beta} K^{T_\alpha\bar{T}_\beta}\left(a_\alpha A_\alpha e^{-a_\alpha T_\alpha}\bar{W}\de_{\bar{T}_\beta}K+a_\beta A_\beta e^{-a_\beta \bar{T}_\beta}W\de_{T_\alpha}K\right)\right]+\\
    &+3\hat{\xi}\frac{\hat{\xi}+7\hat{\xi}\mathcal{V}+\mathcal{V}^2}{\left(\mathcal{V}-\hat{\xi}\right)\left(2\mathcal{V}-\hat{\xi}\right)^2}|W|^2+ \dots
    \\
    \coloneqq &\, V_{\text{np}_1}+V_{\text{np}_2}+V_{\alpha'}+\dots
\end{split}
\label{eq:scalebreak-pot}
\end{equation}
where the dots stand for the pieces corresponding to the $G$ fields only and the mixing of $G$ and $T$. Indeed, in the following analysis we can disregard the $G\bar{G}$ contribution since it will lead to a double-warp suppressed term as in Eq.~\eqref{eq:ISDthraxpot}. For a general treatment of the potential, we should add also the cross terms between the K\"ahler moduli and the $G$ fields. However, it was shown in~\cite{Grimm:2004uq} that the K\"ahler metric for these components is proportional to $b^a$. If the VEV of $b^a$ at the minimum is at most order $\epsilon$ as defined in Eq.~\eqref{eq:genepsilon}, they can be neglected in this analysis because they will produce terms of order $\mathcal{O}(\epsilon^2)$. If the VEV of $b^a$ at the minimum is larger than $\epsilon$, they must be considered. 
For the sake of expositions, we will now assume that the VEV of $b^a$ vanishes at the minimum.\footnote{This assumption is generically satisfied when D-terms are included~\cite{Jockers:2005zy,Grimm:2011dj,Long:2014dta}} The vanishing VEV of $b^a$ implies then that the K\"ahler metric is block diagonal. We do not expect that a different VEV for $b^a$ would change the consequences of our discussion. 

The terms that might lift the thraxion mass are those that break the no-scale condition of the potential and the perturbative corrections in $\alpha'$, i.e. $V_{\text{np}_2}+V_{\alpha'}$. 
This allows us to focus only on the terms in Eq.~\eqref{eq:scalebreak-pot}, since there will be no contributions to the potential coming from the mixing between the $G^a$ fields and the K\"ahler moduli $T_\alpha$ once $b^a=0$.
By decomposing the superpotential and the K\"ahler moduli in real and imaginary parts, i.e.
\begin{equation}
   W=\re(W)+i\im(W)=W^R+iW^I\coma T_\alpha=\re(T_\alpha)+i\im(T_\alpha)=T^R_\alpha+iT^I_\alpha\coma
\end{equation}
we rewrite $V_{\text{np}_2}+V_{\alpha'}$ as follows:
\begin{equation}
\begin{split}
    V_{\text{np}_2}+V_{\alpha'}= &\,e^{K}\left[2\sum_{\alpha,\beta}K^{T_\alpha\bar{T}_\beta} W^Ra_\alpha A_\alpha e^{-a_\alpha T^R_\alpha}\de_{\bar{T}_\beta}K\cos\left(a_\alpha T^I_\alpha\right)+\right.\\
    &-2\sum_{\alpha,\beta}K^{T_\alpha\bar{T}_\beta} W^Ia_\alpha A_\alpha e^{-a_\alpha T^R_\alpha}\de_{\bar{T}_\beta}K\sin\left(a_\alpha T^I_\alpha\right)+\\
    &\left.+3\xi \frac{\xi^2+7\xi \mathcal{V}+\mathcal{V}^2}{(\mathcal{V}-\xi)(2\mathcal{V}+\xi)^2}\left(\left(W^R\right)^2+\left(W^I\right)^2\right)\right]\fstop
\end{split}
\label{eq:dangerPot}
\end{equation}
In Section~\ref{sec:structurePotgen}, we will see that Eq.~\eqref{eq:dangerPot} can endanger the double suppression of the thraxion potential found in~\cite{Hebecker:2018yxs} and displayed in~\eqref{eq:ISDthraxpot}. Generically, the superpotential for the thraxions will generate linear terms in the warp factor in presence of K\"ahler moduli. 
However, we propose two ways in which such situation does not occur and the six-time warp suppression is recovered. In Section~\ref{sec:structurePotdt} we show that by allowing for tuning of fluxes and topological properties of the CY, the linear term in $\epsilon$ vanishes. In Section~\ref{sec:MS-LVS} we comment on how an exponentially large CY volume stabilized à la LVS could compete with $\epsilon$. For particular cases, this makes the linear term in $\epsilon$ subdominant with respect to the $\epsilon^2$ ones.

\subsection{General structure of the superpotential in the presence of thraxions}
\label{sec:structurePotgen}

In this section, we argue that the potential in Eq.~\eqref{eq:dangerPot} actually generates cross terms that are linear in the warp factor $\epsilon\sim \omega_{\IR}^3$. We consider a setup of $n$ multi-throats, each one hosting a number $m_k$ of thraxions, $k=1,\ldots n$. The superpotential reads:\footnote{In this section we use $\mathfrak{I}$ as index to count the number of thraxions, because we reserve $I$ to be the index indicating the imaginary part of a complex function.}
\begin{equation}\label{eq:Wthraxnonpert}
    W_{\text{eff}}=\hat{W}_0-\sum_{k=1}^n\epsilon_{k}e^{i\sum_{\mathfrak{I}=1}^{m_k}p_\mathfrak{I}^kG^\mathfrak{I}/M_k}+\sum_\alpha A_\alpha e^{-a_\alpha T_\alpha}\fstop
\end{equation}
It is possible to divide the superpotential in real and imaginary part, defining
\begin{equation}
\begin{split}
    \hat{W}_0=\re(\hat{W}_0)+i\im(\hat{W}_0)=\hat{W}^R_0+i \hat{W}^I_0\coma \epsilon_k=\re(\epsilon_k)+i\im(\epsilon_k)=\epsilon^R_k+i\epsilon^I_k\coma \\
    G^\mathfrak{I}=\re(G^\mathfrak{I})+i\im(G^\mathfrak{I})=G^\mathfrak{I}_R+iG^\mathfrak{I}_I\coma T_\alpha=\re(T_\alpha)+i\im(T_\alpha)=T^R_\alpha+iT^I_\alpha\coma
\end{split}
\end{equation}
so that
\begin{equation}
    \begin{split}
        W=&\, \hat{W}^R_0+\sum_\alpha A_\alpha e^{-a_\alpha T^R_\alpha}\cos\left(a_\alpha T^I_\alpha\right)-\sum_{k}\epsilon^R_ke^{-\sum_{\mathfrak{I}}p^k_\mathfrak{I}\frac{G_I^\mathfrak{I}}{M_k}}\cos\left(\sum_\mathfrak{I}p^k_\mathfrak{I}\frac{G_R^\mathfrak{I}}{M_k}\right)+\\
        &+\sum_{k}\epsilon^I_ke^{-\sum_{\mathfrak{I}}p^k_\mathfrak{I}\frac{G_I^\mathfrak{I}}{M_k}}\sin\left(\sum_\mathfrak{I} p^k_\mathfrak{I}\frac{G_R^\mathfrak{I}}{M_k}\right)+\\
        &i\left(\hat{W}^I_0
        -\sum_\alpha A_\alpha e^{-a_\alpha T^R_\alpha}\sin\left(a_\alpha T^I_\alpha\right)-\sum_{k}\epsilon^R_ke^{-\sum_{\mathfrak{I}}p^k_\mathfrak{I}\frac{G_I^\mathfrak{I}}{M_k}}\sin\left(\sum_\mathfrak{I}p^k_\mathfrak{I}\frac{G_R^\mathfrak{I}}{M_k}\right)+\right.\\
        &\left.-\sum_{k}\epsilon^I_ke^{-\sum_{\mathfrak{I}}p^k_\mathfrak{I}\frac{G_I^\mathfrak{I}}{M_k}}\cos\left(\sum_\mathfrak{I} p^k_\mathfrak{I}\frac{G_R^\mathfrak{I}}{M_k}\right)\right)\\
        =&\,W^R+iW^I\fstop
    \end{split}
\label{eq:genWthr}
\end{equation}
We can try to explicitly compute $\epsilon_k$ introducing 
\begin{equation}
\begin{split}
    \tilde{g}_{0,k}&=\re(\tilde{g}_{0,k})+i\im(\tilde{g}_{0,k})=\tilde{g}^k_{R,0}+i\tilde{g}^k_{I,0}\coma\\
    g^{jk}_1 & = \re(g^{jk}_1)+i\im(g^{jk}_1)=g^{jk}_{R,1}+ig^{jk}_{I,1}\coma \\
    g^k_{W,1}& = \re(g^k_{W,1})+i\im(g^k_{W,1})=g^k_{R,W,1}+ig^k_{I,W,1}\fstop
\end{split}
\end{equation}
In this way we can define 
\begin{equation}
    z_{0,k}=|R_{0,k}|e^{i\varphi_{0,k}}\coma
\end{equation}
with
\begin{equation}
    \begin{split}
        |R_{0,k}|&=\exp\left[-2\pi \frac{K^k}{g_sM_k}+\frac{2\pi }{M_k}\left(\sum_j M_j g^{jk}_{I,1}+g^k_{I,W,1}+\tilde{g}_{R,0}^k\frac{\hat{W}_{R,0}}{\mathfrak{a}}+\tilde{g}_{I,0}^k\frac{\hat{W}_{I,0}}{\mathfrak{a}}\right)-1\right]\coma\\
        \varphi_{0,k} &= -\frac{2\pi }{M_k}\left(\sum_j M_j g^{jk}_{R,1}+g^k_{R,W,1}-\tilde{g}_{R,0}^k\frac{\hat{W}_{I,0}}{\mathfrak{a}}+\tilde{g}_{I,0}^k\frac{\hat{W}_{R,0}}{\mathfrak{a}}\right)\fstop
    \end{split}
\end{equation}
Finally, $\epsilon_k$ becomes
\begin{equation}\label{eq:epsilonRI}
\begin{split}
    \epsilon_k=&\,\epsilon^R_k+i\epsilon^I_k\\
    = &\,\frac{M_k}{2\pi}|R_{0,k}|\sin(\varphi_{0,k})-\frac{1}{\mathfrak{a}}|R_{0,k}|\left(\tilde{g}^k_{R,0}\hat{W}_0^R+\tilde{g}^k_{I,0}\hat{W}_0^I\right)\sin(\varphi_{0,k})+\\
    &-\frac{1}{\mathfrak{a}}|R_{0,k}|\left(\tilde{g}^k_{R,0}\hat{W}_0^I-\tilde{g}^k_{I,0}\hat{W}_0^R\right)\cos(\varphi_{0,k})+\\
    &+i\left(-\frac{M_k}{2\pi}|R_{0,k}|\cos(\varphi_{0,k})+\frac{1}{\mathfrak{a}}|R_{0,k}|\left(\tilde{g}^k_{R,0}\hat{W}_0^R+\tilde{g}^k_{I,0}\hat{W}_0^I\right)\cos(\varphi_{0,k})+\right.\\
    &\left.-\frac{1}{\mathfrak{a}}|R_{0,k}|\left(\tilde{g}^k_{R,0}\hat{W}_0^I-\tilde{g}^k_{I,0}\hat{W}_0^R\right)\sin(\varphi_{0,k})\right)\fstop
\end{split}
\end{equation}
Assuming that the c.s. moduli stabilization in the bulk is done at energies high enough that do not interfere with the stabilization of the K\"ahler moduli, we can take the functions $g$ in~\eqref{eq:epsilonRI} to be approximately zeros except for $g_{W,0}$ and $g_{K,0}$. This approximation was done in Section 3.2.3 of~\cite{Hebecker:2018yxs}, however, in Section~\ref{sec:structurePotdt}, we show that this assumption is necessary, together with other assumptions, in order to get the six-times suppression of the thraxion masses. As a consequence, $\epsilon_k^R$ in~\eqref{eq:epsilonRI} vanishes.

Another possible assumption is that $\hat{W}_0$ is purely real,\footnote{This assumption is not strictly necessary, but it could be another way to simplify the expression. Note that $W_0$ from the c.s. moduli stabilization is related to $\hat{W}_0$ by a shift of a function depending on $\epsilon_k$. The requirement that $\hat{W}_0$ is completely real, means that we are allowing for small imaginary parts for $W_0$.} so that, finally, 
\eqref{eq:genWthr} becomes
\begin{equation}
    \begin{split}
        W=&\, \hat{W}^R_0+\sum_\alpha A_\alpha e^{-a_\alpha T^R_\alpha}\cos\left(a_\alpha T^I_\alpha\right)+\sum_{k}\epsilon^I_ke^{-\sum_{\mathfrak{I}}p^k_\mathfrak{I}\frac{G_I^\mathfrak{I}}{M_k}}\sin\left(\sum_\mathfrak{I} p^k_\mathfrak{I}\frac{G_R^\mathfrak{I}}{M_k}\right)+\\
        &-i\left(
        \sum_\alpha A_\alpha e^{-a_\alpha T^R_\alpha}\sin\left(a_\alpha T^I_\alpha\right)+\sum_{k}\epsilon^I_ke^{-\sum_{\mathfrak{I}}p^k_\mathfrak{I}\frac{G_I^\mathfrak{I}}{M_k}}\cos\left(\sum_\mathfrak{I} p^k_\mathfrak{I}\frac{G_R^\mathfrak{I}}{M_k}\right)\right)\fstop
    \end{split}
    \label{eq:genWsimpl}
\end{equation}
Recall that in Eq.~\eqref{eq:dangerPot}, we have ignored the cross terms proportional to $b^a$ because we will evaluate the potential at the minimum. We notice that a linear term in $\epsilon$ survives in the scalar potential. Because of this linear dependence of $V$ in $\epsilon$ we expect that generically the thraxion mass scales linearly with the warp factor.

One could naively argue that, in the case in which the minimum is realized at $T^I_{\alpha}=\kappa\pi/a_{\alpha}$, such linear dependence of $V$ in $\epsilon$ vanishes, as in this case the whole second line of equation~\eqref{eq:dangerPot} vanishes.\footnote{Moreover, the piece proportional to $(W^I)^2$ also will not have a linear piece in $\epsilon$.} However, we remark that this is not the case, as the $W^R$ term in~\eqref{eq:genWsimpl} will still carry a linear dependence in $\epsilon$. Moreover, we generically do not expect $T^I$ to stabilize at such VEV. Despite this, we will see in the next section that in some specific models the opposite is true and $T^I$ stabilizes at zero. We will expand on this point in the next sections.

\subsection{Vanishing conditions of the $\mathcal{O}(\epsilon)$ cross terms and application to KKLT}
\label{sec:structurePotdt}

We showed that \emph{generically}, the thraxion potential receives non-trivial contributions of order $\mathcal{O}(\epsilon)$ from K\"ahler moduli stabilization, which spoil their characteristic six-time-warp suppressed scale. However, in some cases, such contributions to the scalar potential can vanish. In this section we first perform the KKLT moduli stabilization procedure with a simplified thraxion superpotential. Then, we comment on some possible ways to cancel the terms of the potential which are linear in $\epsilon$.

Consider a setup of $n$ multi-throats, each one hosting $m_k$ thraxions, $k=1,\ldots n$. Suppose that the $k$-th multi-throat has $n_k$ interpolating $\mathcal{A}$-cycles and $\mathcal{B}$-cycles. 
Then, we allow the following simplifications:
\begin{itemize}
    \item $m_k=1, \forall k=1, \ldots n$.
    \item For each multi-throat system, the flux quanta are chosen with the same magnitude, i.e. $|K_{i,k}|=c_k, \ |M_{i,k}|=d_k, \  \forall\, i=1,\ldots n_k$ and given fixed integer numbers $c_k, \ d_k$.
    \item The homology relation defining the single thraxion present in the multi-throat is of the form $\sum_j [\mathcal{A}_j]=0$, namely $p_j^k=1, \ \forall j, k$.
   \item All $\epsilon_k$ are equal. 
\end{itemize}
It is straightforward to show that under these assumptions the  superpotential~\eqref{eq:thraxionsuperpot} in a single multi-throat system plus the non-perturbative corrections takes the form
\begin{equation}\label{eq:dtsuperpotnp}
    W= W_0+\epsilon\left(1-\cos\left(\frac{G}{M}\right)\right)+\sum_{\alpha}A_\alpha e^{-a_\alpha T_\alpha}\coma
\end{equation}
where $\epsilon=\epsilon^R+i\epsilon^I$.\footnote{Notice that we reabsorbed the factor proportional to $n$ in the definition of $\epsilon$.} 
The KKLT scenario achieves K\"ahler moduli stabilization by including only the non-perturbative corrections to the superpotential. All K\"ahler moduli are stabilized to a SUSY AdS minimum. A necessary condition for the KKLT scheme to hold is that the c.s. stabilization is performed such that $W_0$ is very small. This is needed in order to stabilize at large volume and ignore possible corrections. Lately, this has been proven to be achievable in a series of controlled setups~\cite{Demirtas:2019sip,Demirtas:2020ffz,Alvarez-Garcia:2020pxd}. In order to include the thraxion in the KKLT scenario, we use the superpotential of Eq.~\eqref{eq:dtsuperpotnp}.
As a concrete example, we consider the stabilization of one K\"ahler modulus in presence of one thraxion. The K\"ahler potential is
\begin{equation}
K_{\text{thr}}=-3 \log\left(F\right)\coma \mbox{where }\,
		F= T+\bar{T} -\frac{g_s }{4}\kappa_{+--}\left(G-\bar{G}\right)^2\fstop
\end{equation}
After stabilizing $b$ to zero, the F-term scalar potential reads~\cite{Moritz:2019dxj}
  \begin{equation}\label{eq:KKLTpotential}
  	\begin{split}
  		e^{-K_{\text{c.s.}}}\cdot V=&\,\frac{a A^2 e^{-2 a T^R}\left(aT^R+3\right)}{6(T^R)^2} - \frac{|\epsilon|^2}{6 (T^R)^2 M^2 g_s \kappa_{+--}}\sin\left(\frac{c}{M}\right)^2+\\
  		& +\frac{a A }{2(T^R)^2}\re\left[\bar{W_0}e^{-a T} +\bar{\epsilon}\,e^{-a T}\left(1-\cos \frac{c}{M}\right)\right]\fstop
  	\end{split}
  \end{equation}
We have some new pieces compared to the usual $V_{\text{KKLT}}$ without thraxions. The first one is the $G\bar{G}$ term found in \eqref{eq:ISDthraxpot}: it scales as $\epsilon^2\sim \omega_{\IR}^6$. Instead, the cross term is new, it is induced by the presence of no-scale breaking effects and it scales as $\epsilon\sim \omega_{\IR}^3$. Thus, even in the easiest toy model, we obtain a term which lifts the double suppression of the thraxion mass to a single suppression.

Now, we investigate if we can remove the cross term and restore the six-time warp suppression. Consider adding to the bullet list above the additional requirement:
\begin{itemize}
    \item All $\epsilon_k$ are imaginary, i.e. $\epsilon_k^R=0$. 
\end{itemize}
Hence, we see already from the toy model that with this additional request, the cross term cancels when the $C_4$ axion is stabilized to its minimum. In general, we can show this process as follows. We can expand~\eqref{eq:dtsuperpotnp} with $\epsilon^R=0$ in its real and imaginary parts as
\begin{equation}
    \begin{split}
        W=&\,W_0+\sum_{\beta}A_\beta e^{-a_\beta T^R_\beta}\cos\left(a_\beta T^I_\beta\right)+\epsilon^I\sin\!\left(\!\frac{c}{M}\!\right)\sinh\!\left(\!\frac{b}{g_sM}\!\right)+\\
        &-i\left(\sum_{\beta}A_\beta e^{-a_\beta T^R_\beta}\sin\left(a_\beta T^I_\beta\right)+\epsilon^I\left(\!\cos\!\left(\!\frac{c}{M}\!\right)\cosh\!\left(\!\frac{b}{g_sM}\!\right)-1\right)\right)\\
        =&\,W^R+i W^I\fstop
    \end{split}
\end{equation}
First, we see that, when $b = 0$, $W^R$ does not contain $\epsilon^I$, so the cross terms that were present in Section~\ref{sec:structurePotgen} cancel out. Moreover, it is possible to see that $T^I_\beta$ stabilizes at $\kappa\pi/a_\beta$, with $\kappa\in \ZZ$. The only terms that contain $\epsilon^I$ are those multiplied by $W^I$. However, they cancel when the potential is evaluated at $T^I=\kappa \pi/a_\beta$. All the terms that could possibly give cross terms are then canceled and the final potential for the thraxion scales as in Eq.~\eqref{eq:ISDthraxpot}, i.e. with the six-time warp factor.

Many moduli stabilization scenarios naturally minimize at $b=0$. 
However, stabilizing the $b$ field to zero carries about another important consequence that could help to reduce the amount of tuning required to cancel the linear terms in $\epsilon$. As shown in~\cite{Hebecker:2018yxs}, a non-vanishing VEV for the $b$ field produces a backreaction on the throats as it changes the relative $H_3$-flux distribution. In other words, this makes all the throats (in the same multi-throat system) of different lengths. In this case, the warp factors, i.e. the $\epsilon$ parameters, acquire all different values. In turn, this means that in the case in which $b=0$, all the warp factors in the same multi-throat system could be taken to be equal more easily. As $\epsilon_{k}\sim e^{-K_k/g_s M_k}$, one still has to require that the ratio of the flux numbers is equal in each throat of the system. Once these two requirements are met, all the warp factors in the same multi-throat system are actually equal. We note here that in a double-throat system, this is always the case. Thanks to the homology relation, we have only one effective $\mathcal{A}$- and $\mathcal{B}$-cycle, hence only one effective flux quantum of $F_3$ and of $H_3$.  

We comment now on the conditions listed above. The  first condition states that in every multi-throat system there must be a single homology relation and therefore a single thraxion. It is interesting to notice that in \emph{all} known examples of CY orientifold supporting thraxions, this is always true. Namely, at the moment of writing we do not know of any CY orientifold in which a given multi-throat hosts more than one thraxion. However, in view of how small the set of CICY parents of our CICY orientifold database is compared to other known algorithmically constructable sets of CY 3-folds, it seems unwarranted to assume a priori that manifolds with multi-throats hosting more than one thraxion do not exist. We comment more on this in Section~\ref{sec:MultiThrax-database}. 

The second condition implies that all the various throats in the same multi-throat system have the same length. The third condition requires a specific form of the homology relation. Notice that one could use a rescaling of the base of 3-cycles $[\mathcal{A}_j]\to n_j[\mathcal{A}_j]$, $[\mathcal{B}_j]\to n_j^{-1}[\mathcal{B}_j], \ n_j\in \mathbb{Z}\setminus\{0\}$ in order to ensure that such condition is always satisfied. We remark that if we have more than one homology relation, only one of them can be recast in the form $\sum_j[\mathcal{A}_j]=0$ by rescaling. Hence, for multi-throats carrying more than one thraxion the symmetrization of the multi-throat becomes impossible.

The last condition, $\epsilon^R_k=0$, is observed to restore the six-times warp suppression of the thraxion mass. Moreover, for double-throats (provided stabilizing $b=0$) ensuring $\epsilon^R_k=0$ guarantees the enhanced warp suppression of the thraxion mass. Hence, it is interesting to note that in the subclass of flux vacua found in~\cite{Demirtas:2020ffz}, which is determined by the prime condition necessary for well working KKLT vacua (i.e. small $W_0$), $\epsilon$ is always imaginary to leading order in the conifold modulus $z_0$.\footnote{We thank J. Moritz for pointing this out to us.} This should not be seen as a physical motivation, but rather as evidence supporting the existence of whole classes of examples realizing this assumption. 

In this section, we argued that under some special conditions, the thraxion mass can still be double-suppressed. However, these requirements are generically difficult to meet in a more complicated scenario in which within a given multi-throat system there is more than one thraxion, or unequal flux ratios.

\subsection{Behavior of the ${\cal O}(\epsilon)$ thraxion mass cross terms in LVS}
\label{sec:MS-LVS}
There is another, interesting way which could restore the six-times warp suppression. Such way appears to be quite generic as long as one stabilizes the CY volume $\mathcal{V}$ to exponentially large values, as happens in LVS. In the following, we show how the interplay between large values of $\mathcal{V}$ and small values of $\epsilon$ could favor the terms proportional to $\epsilon^2$ over the linear ones.

LVS stabilizes the K\"ahler moduli via an interplay between $\alpha'$ corrections to $K$ and non-perturbative corrections to $W$. In order to keep control on $\alpha'$ corrections, the overall CY volume $\mathcal{V}$ must be stabilized at exponentially large values. As a result, $W_0$ takes $\mathcal{O}(1)$ values. However, some topological requirements are needed. First, inside the $\alpha'$ correction in~\eqref{eq:KpotBBHL}, the Euler number of the CY must be negative, i.e. $h_{2,1} >h_{1,1}>1$. This in turn ensures that $\xi$ is positive and so that, as $\mathcal{V}\to \infty$, the potential goes to zero from below. LVS produces an AdS minimum which is no longer supersymmetric. Second, there must be present at least one blow-up mode, $\tau_s\subset T_s$, corresponding to a 4-cycle modulus resolving a pointlike singularity. In the limit $\mathcal{V}\to \infty$, all $\tau_i\to \infty$ but $\tau_s$. This modulus should be the one inducing the leading non-perturbative corrections to $W$, namely $W_{\text{np}}\sim e^{- a_s T_s}$. Of course all moduli could appear in $W_{\text{np}}$, but in the above limit their contribution is subleading. 

More in detail, by considering a superpotential corrected with~\eqref{eq:npsuperpot} together with the K\"ahler potential in~\eqref{eq:KpotBBHL}, LVS stabilizes the volume as $\mathcal{V}\sim e^{a_s T^R_s}$. In turn, this means that in the potential, each time a term is proportional to $e^{-n\, a_s T^R_s}$, such term is $\mathcal{O}(\mathcal{V}^{-n})$ times suppressed. The standard LVS potential without the odd sector scales as $\mathcal{O}(\mathcal{V}^{-3})$ \cite{Balasubramanian:2005zx}.\footnote{See also~\cite{Gao:2013rra} for the inclusion of the odd sector in LVS, where the odd axions get a potential from fluxed D3-brane instanton contributions to $W$, as well as~\cite{Cicoli:2021tzt} for very recent results on moduli stabilization with odd axions.} Let us now consider the superpotential in Eq.~\eqref{eq:Wthraxnonpert}. For $\mathcal{V}\to \infty$, the no-scale breaking potential of Eq.~\eqref{eq:scalebreak-pot} scales as\footnote{Here and below we make use of the no-scale breaking property of the K\"ahler potential and of the following standard relations about the K\"ahler metric, as derived in~\cite{Grimm:2004uq}, in the $\mathcal{V}\to \infty$ limit: $$K_{T_\alpha}\sim t^\alpha \mathcal{V}^{-1}\coma  K^{T_\alpha \bar{T}_\beta}\sim -\mathcal{V} \kappa_{\alpha\beta\gamma} t^\gamma + \tau_\alpha\tau_\beta\coma  K^{T_\alpha \bar{T}_\beta}K_{T_\alpha}\sim -\tau_\beta \coma K^{G^a \bar{G}^b}\sim -g_s^{-1} \mathcal{V} \left(\kappa_{ab\gamma}t^\gamma\right)^{-1} 
\fstop $$ Since at this stage we are only interested in the scaling with $\mathcal{V}$, we will drop numerical factors, the signs and the dependence on the 2-cycles. We will restore them when computing the explicit example.}
\begin{equation}\label{eq:potLVSscaling}
    \begin{split}
        V_{\text{np}_1}&\sim \frac{K^{T_s \bar{T}_s}|\partial_{T_s}W|^2}{\mathcal{V}^2} \sim   \mathcal{O}\left(\frac{1}{\mathcal{V}^3}\right)+ \dots  \\
        V_{\text{np}_2}&\sim - \frac{K^{T_s \bar{T}_s}K_{\bar{T}_s}}{\mathcal{V}^2}\,W^R e^{-a_s T_s^R}\cos\left(a_s T_s^I\right)- \frac{K^{T_s \bar{T}_a}K_{\bar{T}_a}}{\mathcal{V}^2} e^{-a_s T_s} \bar{W} \\
        &\sim \mathcal{O}\left(\frac{1}{\mathcal{V}^3}\right)+\mathcal{O}\left(\frac{\epsilon}{\mathcal{V}^3}\right)+\dots \\ 
        V_{\alpha'}&\sim \frac{\hat{\xi} |W|^2}{\mathcal{V}^3} \sim \mathcal{O}\left(\frac{1}{\mathcal{V}^3}\right)+ \mathcal{O}\left(\frac{\epsilon}{\mathcal{V}^3}\right)+ \dots \fstop
    \end{split}
\end{equation}
However, the $G\bar G$ part of the potential has a different scaling, namely 
\begin{equation}\label{eq:LVSGGbarpot}
    V_{G,\bar G}=  K^{G\bar G} \mathcal{D}_G W\mathcal{D}_{\bar G}\bar W =\frac{K^{G\bar G}|\partial_G W|^2}{\mathcal{V}^2}\sim \mathcal{O}\left(\frac{\epsilon^2}{\mathcal{V}}\right)\fstop
\end{equation}
This piece receives volume-suppression only from $e^K$, which is partially compensated by the inverse of the K\"ahler metric $K^{G\bar G}$. This results in an $\mathcal{O}(\mathcal{V}^{-1})$ suppression, which is milder than the $\mathcal{O}(\mathcal{V}^{-3})$ dependence of the term proportional to $\epsilon$ in the potential~\eqref{eq:potLVSscaling}. Therefore, in LVS the stronger suppression in $\epsilon^2$ is compensated by a milder one in $\mathcal{V}$ and hence it could happen that the $\mathcal{O}(\epsilon^2)$ term coming from~\eqref{eq:LVSGGbarpot} would dominate over the $\mathcal{O}(\epsilon)$ one. Notice that, so far, the discussion is completely general.

In the following, we show this remarkable behavior in a specific example. Then, we comment on the implications of the interplay between $\epsilon$ and $\mathcal{V}$ in two phenomenological applications. For the sake of consistency, we explicitly compute the F-term scalar potential for a CY with one thraxion, $h_+^{1,1}=2$ and whose volume takes the standard Swiss-cheese form
\begin{equation}
 		\mathcal{V}=\left(T_b+\bar{T}_b\right)^{3/2} - \left(T_s+\bar{T}_s-\frac{g_s}{4}\kappa_{s--}\left(G-\bar{G}\right)^2\right)^{3/2}
 		\coma
 \end{equation}
 where we assumed that the only nontrivial even-odd-odd triple intersection number is $\kappa_{s--}$. In Appendix~\ref{app:LVS} we show that considering all couplings to be nontrivial leads to a potential with the same structure. We further assume that the thraxion superpotential can take the form of Eq.~\eqref{eq:1dtsuperpot} and hence the total superpotential for this toy model can be written as
  \begin{equation}
 	W(G,T_s)=W_0+ \epsilon \left(1-\cos\left(G/M\right)\right) +A_s e^{-a_s T_s}\coma
 \end{equation}
 where the leading non-perturbative correction comes from the blow-up modulus $\tau_s$. The field $b$ stabilizes at zero, and in order for its kinetic terms to be positive definite we should have $\kappa_{s--}>0$. Hence, we get the following potential
 \begin{equation}
 	\begin{split}
 	e^{-K_{\text{c.s.}}} V=\,& \frac{2\sqrt{2}a_s^2 A_s^2\, e^{-2 a_s \tau_s}\sqrt{\tau_s}}{3\mathcal{V}}+\frac{4 a_s A_s W_0\tau_s e^{-a_s \tau_s}\cos\left(a_s\theta_s\right)}{\mathcal{V}^2}+\frac{3W_0^2\hat{\xi}}{4\mathcal{V}^3}+\\
 	&-\frac{3 \epsilon^2\,\hat{\xi}}{\mathcal{V}^3}\sin\left(\frac{c}{2M}\right)^4-\frac{\epsilon^2\sqrt{2}\left(4\mathcal{V}^2-2\mathcal{V}\hat{\xi}+\hat{\xi}^2\right)}{12 g_s M^2 \kappa_{s--}\sqrt{\tau_s}\,\mathcal{V}^3}\sin\left(\frac{c}{M}\right)^2+\\&+\frac{4 i\, \epsilon\, a_s A_s \tau_s \, e^{-a_s\tau_s}}{\mathcal{V}^2}\left(1-\cos\left(\frac{c}{M}\right)\right)\sin\left(a_s \theta_s\right)\\
 	\equiv & \,V_{\text{LVS}}\left(\mathcal{V},\tau_s,\theta_s\right)+V_{\text{thr}}\left(\mathcal{V},\tau_s,\theta_s,c\right)\fstop
 \end{split}\label{eq:LVSpot21}
 \end{equation}
Notice that for $\epsilon=0$ we recover the standard LVS potential, and the thraxion $c$ enters as a correction in $\mathcal{O}(\epsilon)$ and $\mathcal{O}(\epsilon^2)$. Then, the LVS moduli stabilization proceeds as usual. Also, this turns out to be one of the special cases in which the thraxion is independent of the stabilization of the K\"ahler moduli, as the $\mathcal{O}(\epsilon)$ term vanishes once the $C_4$ axion is set in its VEV.
However, our main point now is the following:  in~\eqref{eq:LVSpot21} the $\mathcal{O}(\epsilon)$ term is twice more suppressed in $\mathcal{V}$ than one of the $\mathcal{O}(\epsilon^2)$ ones. Therefore, it could happen that these effects balance among each other and the $\mathcal{O}(\epsilon^2)$ term becomes eventually the leading one. 

As a first application, we could investigate whether this situation takes place when we require the thraxion to be the inflaton. Suppose we want to realize an inflationary potential with $V\sim (10^{15}\text{ GeV})^4$, i.e. $V\sim 10^{-12}$ in Planck units. In order for the $\mathcal{O}(\epsilon^2)$ term to be the leading one and thus to reproduce such scaling, we should have $\mathcal{V}>250$. This guarantees that the $\mathcal{O}(\epsilon)$ term is subleading. The value we found for the CY volume fits perfectly within LVS. Therefore, for inflationary applications, we restore the double suppression of the thraxion mass as in its original proposal.

Nevertheless, if we consider the thraxion to be a possible Fuzzy Dark Matter (FDM) candidate as in~\cite{Cicoli:2021gss}, this balance turns out to be impossible, or very dangerous for LVS. With FDM we refer to a particle taken as a possible dark matter candidate that is so light that its nature is basically wave-like~\cite{Hu:2000ke}. Such particle should be characterized by a mass of order $m\sim 10^{-22}$ eV and a decay constant of roughly $f\sim 10^{17}$ GeV. In~\cite{Hui:2016ltb}, it was shown that for a stringy axion-like particle to be a good FDM candidate, its instanton action (and thus the potential) should scale as $e^{-S}\sim e^{-230}\sim 10^{-100}$ in Planck units. Requiring the $\mathcal{O}(\epsilon^2)$ term to be the leading one entails a large size for the CY volume, namely $\mathcal{V}>10^{20}$. However, such value is incompatible with the low energy phenomenology. Given that the scale of SUSY breaking is $m_{3/2}\sim \mathcal{V}^{-1}M_{P}$, we would have SUSY at values smaller than $10^{-2}$ GeV. Therefore, for FDM in a LVS moduli stabilization, the leading term is always the $\mathcal{O}(\epsilon)$ one (if it does not get canceled by the $C_4$ axion stabilization).

\subsection{Mass scales for thraxion setups in KKLT and LVS}

We can now apply our results to derive the mass scaling of the low-lying states in setups with volume moduli stabilization. These light states include the lightest K\"ahler moduli, the warped KK modes inside the multi-throat carrying the thraxion, and the thraxion itself. 

\subsubsection{KKLT}

We begin with the KKLT scenario. Fluxes inside a warped throat generically induce perturbations which scales with powers of $r$. These perturbations are divided in normalizable and non-normalizable modes. In particular, non-normalizable modes correlate with the ISD breaking fluxes~\cite{Gandhi:2011id}.  Since gaugino condensation (necessary to stabilize K\"ahler moduli) breaks both no-scale and sources non-ISD fluxes~\cite{Baumann:2010sx}, its presence activates the non-normalizable perturbation.
 Following the classification of throat perturbations in~\cite{Baumann:2010sx}, we then have the gaugino condensate sourcing a perturbation
\begin{equation}
    \delta G_3 \sim \langle\lambda\lambda\rangle\, r^{-3/2} \sim W_0\, r^{-3/2}\fstop
\end{equation}
As discussed in~\cite[Section 2.4]{Baumann:2010sx} we have to require the coefficients of such perturbations in the UV to be small enough such that at the IR end of the throat they do not become comparable with the background fluxes. In our case we see that the $\langle\lambda\lambda\rangle$-sourced non-ISD flux perturbation becomes $\mathcal{O}(1)$ whenever $r^{3/2}\sim W_0$. Hence, for our perturbation to satisfy the condition of~\cite{Baumann:2010sx} we must ask for\footnote{We thank J. Moritz for recalling to our attention the discussion and condition in~\cite{Baumann:2010sx} and its implication for the IR warp factor.}
\begin{equation}
    \epsilon > W_0^2 \fstop
    \label{eq:a0scalingW0}
\end{equation}
Let us now use Eq.~\eqref{eq:a0scalingW0} to compare the mass of the thraxions to the masses of the other light particles in the compactification. In this section we will focus on KKLT scenarios, while in the following section we will discuss similar computations for LVS.

From the potential in Eq.~\eqref{eq:KKLTpotential}, we have that
\begin{equation}
    m^2_{\text{thr}}\sim \epsilon\, \frac{|W_0|}{\mathcal{V}^{4/3}}\sim m^2_{\text{wKK}}\,\epsilon^{1/3}|W_0|\coma
\end{equation}
where we used the definition $m^2_{\text{wKK}}\sim \epsilon^{2/3}/\mathcal{V}^{4/3}$ for the mass squared of the warped KK modes. Since $|W_0|\ll 1$ in KKLT and $\epsilon\leq1$ by definition, this implies that thraxions stay parametrically lighter than the warped KK modes even if the cross term lifts the thraxion mass-squared to ${\cal O}(\epsilon)$. Moreover, the condition~\eqref{eq:a0scalingW0} implies that the ratio
\begin{equation}
\frac{m^2_{\text{thr}}}{m^2_{\text{wKK}}}> |W_0|^{5/3}
\end{equation}
is bounded from below.

Then, we should compare the mass of K\"ahler moduli with the one of thraxions, to ensure that the latter are still the lightest particle in the spectrum. The K\"ahler modulus mass reads
\begin{equation}
    m_\tau^2 \sim \frac{|W_0|^2}{\mathcal{V}^2}\fstop
\end{equation}
Hence
\begin{equation}
    \frac{m_{\text{thr}}^2}{m_\tau^2}\sim \frac{\mathcal{V}^{2/3}\epsilon}{|W_0|}\sim \epsilon\,\frac{\log\left(W_0^{-1}\right)}{|W_0|}\coma
\end{equation}
where we used the KKLT relation $\mathcal{V}\sim - \log(|W_0|)^{3/2}$ for the CY volume (in case of single modulus). We see that $m_{\text{thr}}^2<m_\tau^2$ if $\epsilon<|W_0|$. Therefore, to ensure that the thraxion is lighter than the K\"ahler modulus and that the IR end of the throat is safe from large corrections, $\epsilon$ should sit in the window $|W_0|^2<\epsilon<|W_0|$. 

We can also consider the scaling of the gravitino mass $m_{3/2}\sim |W_0||\log(W_0)|^{-3/2}$ in Planck units, then we get the relation
\begin{equation}
    \frac{m_\tau^2}{m_{\text{thr}}^2}\sim \frac{m_{3/2}}{\epsilon}\,|\log W_0|^{1/2}\fstop
\end{equation}
Therefore, the K\"ahler modulus is heavier compared to the thraxion if
\begin{equation}
    m_\tau^2 > m_{\text{thr}}^2 \quad\Longleftrightarrow \quad\frac{m_{3/2}}{M_P}> \frac{\epsilon}{\sqrt{|\log W_0}|}\coma
\end{equation}
where in the last relation we restored the Planck mass. By requiring that $m_\tau^2 > m_{\text{thr}}^2$ together with the relation~\eqref{eq:a0scalingW0}, we have a lower bound on the gravitino mass.

\subsubsection{LVS}
In LVS, we have that the thraxion is always lighter than the warped KK modes, as
\begin{equation}
 \frac{m_{\text{thr}}^2 }{m_{\text{wKK}}^2}\sim \frac{\epsilon}{\mathcal{V}^{5/3}}\fstop  
 \end{equation}
Then, we can compare the thraxion mass to the one of the volume-supporting K\"ahler modulus, as it is the lightest modulus in the LVS spectrum. The mass squared of the big cycle scales as $m_{\tau_b}^2\sim \mathcal{V}^{-3}$. We see that
\begin{equation}
 \frac{m_{\text{thr}}^2 }{m_{\tau_b}^2}\sim \epsilon \coma
 \end{equation}
which means that the thraxion is always lighter. 

We note here that both the non-perturbative effect stabilizing $\tau_s$ and the ${\cal O}(\alpha'^3)$ correction inferred from the 10d $R^4$ term via the induced correction to the volume moduli K\"ahler potential break no-scale as well as likely source non-ISD 3-form fluxes. By an analysis similar to the one in the KKLT section above, this will source perturbations in the thraxion multi-throat which will bound $\epsilon$ from below. However, doing this properly while including the perturbations sourced by the ${\cal O}(\alpha'^3)$ correction to $K$ is difficult, as the structure of the direct 10d origin, schematically represented by terms $\alpha'^3 G_3^2 R^3$ is unknown. Hence, we have to leave a proper analysis for a future time when the corresponding supersymmetric completion of the $R^4$ term in type IIB string theory will have been determined.

\subsection{Comments about 10d origin of the cross terms}
\label{sec:10Dcross}

We will review here shortly an argument given in~\cite[Section 6.8]{Moritz:2019dxj} concerning the $10$d origin of the single-warped cross term in the thraxion scalar potential with moduli stabilization in many cases. The starting point is the observation that as soon as non-perturbative effects like gaugino condensation on a stack of $4$-cycle wrapping D$7$-branes are involved in K\"ahler moduli stabilization, these non-perturbative effects tend to generate non-ISD $3$-form flux contributions. For simplicity, we now focus on the situation of a double throat. As thraxions are the lowest lying radial KK-mode of $C_2$ in the double throat, they react sensitively not just to the change of the IR Dirichlet boundary conditions driven by giving thraxion a finite VEV, but to changes of the UV boundary conditions as well. The presence of ISD-breaking non-perturbative effects in the bulk in general will source such UV boundary terms which take the form~\cite{Moritz:2019dxj}
\begin{equation}
\delta S[z]=\frac{M_{10}^8}{2}\int d^4x\int \frac{dr}{r}\left(J_{\rm UV}\bar z +c.c.\right)\fstop
\end{equation}
If we use for moduli stabilization e.g. gaugino condensation as the non-perturbative effect, this implies a source $J_{\rm UV}=j \cdot r\,\delta(r-r_{\rm UV})$ with $j \sim \langle\lambda\lambda\rangle$. In the dual holographic description of a perturbed KS throat~\cite{Baumann:2010sx} used to describe each half of the double throat, this corresponds to a dimension $\Delta=3$ chiral operator. In presence of this UV perturbation the solution for $z(r)$ takes the form~\cite{Moritz:2019dxj}
\begin{equation}
z(r) = \frac14 j \, (r^2-r_{\rm IR}^2)+z_1+ \frac 12 \frac{r^2-r_{\rm IR}^2}{r_{\rm UV}^2-r_{\rm IR}^2} (z_2-z_1)
\end{equation}
in the first throat, and with $z_1\leftrightarrow z_2$ in the second throat. If we now insert this back into the $5$d effective action for the complex structure modulus $z(r)$ from~\cite{Hebecker:2018yxs,Moritz:2019dxj} and corrected by $\delta S[z]$, we get a scalar potential
\begin{equation}
\begin{split}
V(c)&=|z_1-z_2|^2+{\rm Re}\,(\bar j(z_1+z_2))+\text{ const.}\nonumber\\
&= 4|z_0|^2-2{\rm Re}\,(\bar j  z_0)(1-\cos(c/M))+\text{ const.}
\end{split}
\end{equation}
which clearly shows the ISD-violating source $j$ generating the cross term, provided that $\bar j  z_0$ is not purely imaginary. Moreover, since e.g. $j\sim\langle\lambda\lambda\rangle\sim |W_0|$ in the KKLT scenario, we see that this argument has the features to reproduce the cross term observed in the 4d EFT computation. We leave a more detailed construction of this argument for future work.

Finally, a full treatment of our $4$d EFT results in a $10$d setting is difficult at the current time, in particular for the case when we choose LVS to stabilize the K\"ahler moduli. The reason here consists of the fact that for LVS, the ${\cal O}(\alpha'^3)$ correction in the volume moduli K\"ahler potential contributes to the ${\cal O}(\epsilon)$ cross term in the scalar potential. However, the contributions to the scalar potential at  ${\cal O}(\alpha'^3)$ were derived from 10d in~\cite{Becker:2002nn} using the known 10d type IIB $\alpha'^3 R^4$ correction dimensionally reducing to the known ${\cal O}(\alpha'^3)$ correction to $K$, from which in turn via 4d ${\cal N}=1$ local SUSY~\cite{Becker:2002nn} inferred the ${\cal O}(\alpha'^3)$ to the supergravity F-term scalar potential. The direct 10d origin of this correction to $V$ would arise from terms reading schematically as ${\cal O}(\alpha'^3)G_3^2 R^3$ which are part of the SUSY completion of the $R^4$ term in type IIB in 10d. This SUSY completion, however, unfortunately is to date not completely known already at the needed fifth order. Hence, at least for the case of LVS K\"ahler moduli stabilization, a 10d discussion of our 4d EFT results must await future progress on the SUSY completion of the type IIB $\alpha'^3 R^4$ term.

\section{Concrete Calabi-Yau Orientifolds Supporting Thraxions}
\label{sec:MultiThrax-database}

In this section, we discuss explicit examples of CY orientifolds which support multi-throat systems hosting thraxions. We work with the set of manifolds known as CICYs~\cite{Candelas:1987kf}. The manifolds in this class are defined as the zero-locus of a set of $k$ homogeneous polynomials $p_j \left(z\right)$ in an ambient space given by $\mathcal{A}=\prod_i \mathbb{P}^{n_i}$, constrained by
\begin{equation}
    \sum_{i}n_i-k=3\fstop
\end{equation}
The multi-degrees of the polynomial equations with respect to the coordinates of the ambient space factors are encoded in a configuration matrix
\begin{equation}
\left[
\begin{tabular}{c|cccc}
$\PP^{n_1}$ &   $q_1^1$ & $\cdots$  & $q_k^1 $ \\
$\PP^{n_2}$  &   $q_1^2$ & $\cdots$  &$ q_k^2$  \\
$\vdots$ &   $\vdots$ & $\ddots$ & $\vdots$  \\ 
$\PP^{n_s} $&   $q_1^s $& $\cdots$ & $q_k^s $
\end{tabular}
\right] \fstop
\label{eq:configuration}
\end{equation}
Requiring the zero-locus of the $p_j$ to be a CY manifold, the vanishing condition for the first Chern class imposes
\begin{equation} \label{eq:CYpolynomialcondgen}
n_i+1=\sum_{j=1}^{k} q_j^i\coma \quad \forall \, i=1,...s \fstop
\end{equation}
Let $X$ be a CICY with ambient space $\mathcal{A}$. If $h^{1,1}(X)=h^{1,1}(\mathcal{A})$ we say that $X$ is a favorable CICY. All CICYs apart from 70 are favorable~\cite{Anderson:2017aux}. In the following, we restrict ourselves to work with favorable CICYs only.

\begin{figure}[!htp]
    \centering
    \begin{subfigure}[t]{0.49\textwidth} 
    \centering
    \begin{scaletikzpicturetowidth}{\textwidth}
		\begin{tikzpicture}[scale=\tikzscale]
		\begin{axis}[
		ybar,
		width=\textwidth,
		bar width=10pt,
		enlarge y limits= 0.1,
		xlabel={$h^{1,1}$},
		ylabel={\# Orient. w FCPs./ \# Orient. $\%$},
		]
		\addplot+[mark=0,draw=primcol!20!black,fill=primcol, opacity=0.5,every node near coord/.style={text=black},every node near coord/.append style={font=\tiny}] plot coordinates{ (1.0,69.23) (2.0,81.98) (3.0,89.87) (4.0,93.89) (5.0,96.10) (6.0,97.79) (7.0,98.46) (8.0,98.81) (9.0,98.88) (10.00,99.08) (11.00,99.01) (12.00,99.20) (13.00,98.91) (14.00,98.92) (15.00,95.06) };
		\end{axis}
		\end{tikzpicture}
		\end{scaletikzpicturetowidth}
		\caption{}
    \end{subfigure}\hfill
    \begin{subfigure}[t]{0.49\textwidth}
    \centering
    \begin{scaletikzpicturetowidth}{\textwidth}
		\begin{tikzpicture}[scale=\tikzscale]
		\begin{axis}[
		ybar,
		width=\textwidth,
		bar width=1pt,
		enlarge y limits= 0.1,
		xlabel={$h^{2,1}$},
		ylabel={\# Orient. w FCPs./ \# Orient. $\%$},
		]
		\addplot+[mark=0,draw=seccol!20!black,fill=seccol, opacity=0.5,every node near coord/.style={text=black},every node near coord/.append style={font=\tiny}] plot coordinates{ (15.00,95.06) (16.00,98.89) (17.00,99.15) (18.00,99.40) (19.00,98.97) (20.00,99.28) (21.00,99.21) (22.00,99.01) (23.00,98.99) (24.00,99.11) (25.00,98.37) (26.00,98.45) (27.00,98.97) (28.00,97.97) (29.00,98.69) (30.00,97.99) (31.00,97.42) (32.00,97.49) (33.00,97.79) (34.00,96.14) (35.00,98.08) (36.00,96.99) (37.00,94.53) (38.00,96.60) (39.00,97.33) (40.00,91.34) (41.00,95.22) (42.00,94.92) (43.00,95.13) (44.00,90.61) (45.00,92.52) (46.00,93.06) (47.00,93.37) (48.00,93.94) (49.00,93.59) (50.00,91.73) (51.00,85.08) (52.00,86.88) (53.00,89.71) (54.00,86.32) (55.00,83.50) (56.00,92.86) (57.00,89.84) (58.00,76.70) (59.00,92.57) (60.00,80.54) (61.00,77.39) (62.00,84.09) (63.00,85.42) (64.00,86.30) (65.00,70.00) (66.00,94.23) (67.00,100.0) (68.00,59.68) (69.00,87.10) (72.00,88.00) (73.00,75.00) (75.00,70.18) (76.00,57.14) (77.00,100.0) (83.00,100.0) (86.00,39.29) (89.00,60.00) (101.0,50.00) };
		\end{axis}
		\end{tikzpicture}
		\end{scaletikzpicturetowidth}
		\caption{}
    \end{subfigure}
    \caption{Percentage of orientifolds admitting frozen conifold points, with respect of the total number of orientifolds, distributed by $h^{1,1}$ and $h^{2,1}$. We remark that we are only analyzing favorable CICYs, therefore the cutoff at $h^{1,1}=15$.}
    \label{fig:FCPsh11h21}
\end{figure}
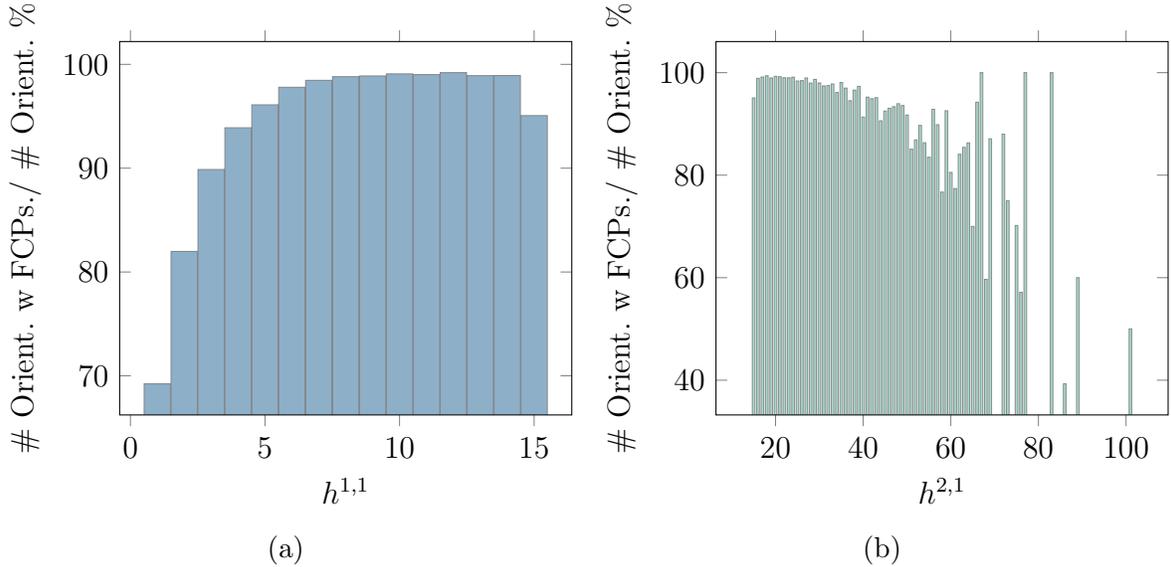

An explicit database listing all the $\ZZ_2$ actions on CICY manifolds which admit fixed loci of codimension 1 and 3, and that descend from involutions of $\mathcal{A}$ was produced in~\cite{Carta:2020ohw}. One important comment is that at a generic point in complex structure moduli space, a CICY will not admit any geometric $\mathbb{Z}_2$ symmetry which could be used in order to define an orientifold projection. However, at special points in c.s. moduli space, such symmetry exists. Due to this complex structure tuning, $\ZZ_2$ symmetric CYs will generically contain conifold singularities that lie on the fixed locus of the $\ZZ_2$ action. Being located on top of an O-plane, these singularities cannot be deformed in a way that is compatible with the $\ZZ_2$ action. However, they can be resolved.\footnote{Just as the usual conifold, they can be resolved in two different ways, related by a flop.} After the orientifold projection, these singularities are called \emph{frozen conifold singularities}. We stress that this feature is extremely generic: in Figure~\ref{fig:FCPsh11h21} we show the percentage of CICY orientifolds for which there exists at least one frozen conifold with respect to all CICY orientifolds, as a function of the Hodge numbers.

We wish now to comment about some aspects of the orientifolds admitting frozen conifolds. From  Figure~\ref{fig:FCPsh11h21} we see clearly that, regardless the presence of thraxions, only ${\cal O}(1)$\% of all orientifolds are free of frozen conifolds. While the CICYs comprise only a comparatively small set of CY manifolds, this outcome raises the possibility that a sizable fraction of all O$7$-orientifolds of CYs may contain frozen conifolds. For example, Table~\ref{tab:percKSfroz} informs us, that even for the 5 single-polynomial CY manifolds which the CICY set has in common with the much larger KS set of anticanonical hypersurfaces in toric ambient spaces~\cite{Kreuzer:2000xy}, about $\sim47\%$ of the possible orientifolds acquire frozen conifolds. The existing partial scans of KS orientifolds, which failed to detect frozen conifolds, may thus be structurally incomplete.

\begin{table}[!htp]
    \centering
    \begin{tabular}{c|c}
        CICY & Orient. w/out FCPs. \\
        \hline
        7890 & $50\%$\\
        7887 & $60\%$\\
        7884 & $0\%$\\
        7880 & $44.4\%$\\
        7862 & $50\%$\\
        \hline
        Total & $46.7\%$
    \end{tabular}
    \caption{Percentage of orientifolds without frozen conifold points for the $5$ CICYs that are also in the KS database.}
    \label{tab:percKSfroz}
\end{table}

If this is the case, this poses the question of understanding in more details the resolution branches of the frozen conifold singularities. This is especially important for phenomenological applications. By entering the resolved phase of a frozen conifold singularity, $h^{1,1}_+$ increases by $\Delta h^{1,1}_{+,\text{f.c.}}$. Thus, new divisors will be present in the resolved phase, compared to the divisors of the double-cover, i.e. the original CY before the orientifold quotient is taken. In turn, this implies that the simple splitting of the $H^{1,1}$(CY)-eigenspace of the parent CY into $\mathbb{Z}_2$-even and odd subspaces to compute the purely even sector and even-odd-odd sector intersection numbers will generically fail to be correct in the resolved phase. Furthermore, such computation is of dubious meaning at the singular point, as the Dolbeault cohomology is not well-defined for singular varieties. Hence, achieving an understanding of the structure and ubiquity of singular CY orientifolds characterized by the presence of frozen conifolds, as well as their resolutions, forms a pressing task for the future.

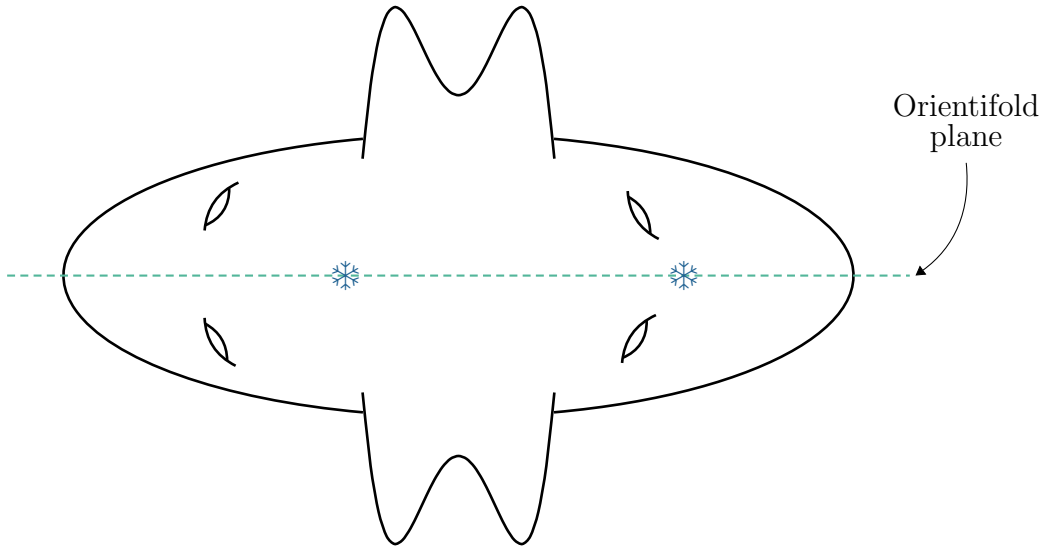
\begin{figure}[!htp]
    \centering
    \begin{tikzpicture}[scale=0.75]
    \begin{scope}[yshift=-1.2cm]
    \draw[scale=1, domain=-1.7:1.7, smooth, variable=\x, line width=1pt,black] plot ({\x}, {\x*\x*\x*\x-2.5*\x*\x});
    \end{scope}
    \draw[line width=1pt] (0,2) [partial ellipse=104:256:7cm and 2.5cm];
     \draw[line width=1pt] (0,2) [partial ellipse=284:436:7cm and 2.5cm];
    \begin{scope}[yshift=5.2cm]
    \draw[scale=1, domain=-1.7:1.7, smooth, variable=\x, line width=1pt,black] plot ({\x}, {-\x*\x*\x*\x+2.5*\x*\x});
    \end{scope}
    \begin{scope}[xshift=2.5cm,scale=0.5,yshift=2cm]
    \draw[line width=1pt] (1.01,4.8) to[bend left=30] (1.8,3.48);
    \draw[line width=1pt] (2.1,3.3) to[bend left] (1,5);
    \end{scope}
    \begin{scope}[xshift=4.25cm,scale=0.5,yshift=-2.9cm]
    \draw[line width=1pt] (-1.83,5.3) to[bend left=35] (-2.69,3.98);
    \draw[line width=1pt] (-2.7,3.8) to[bend left] (-1.5,5.5);
    \end{scope}
    \begin{scope}[xshift=-5cm,scale=0.5,yshift=-2.5cm]
    \draw[line width=1pt] (1.01,4.8) to[bend left=30] (1.8,3.48);
    \draw[line width=1pt] (2.1,3.3) to[bend left] (1,5);
    \end{scope}
    \begin{scope}[xshift=-3.15cm,scale=0.5,yshift=1.8cm]
    \draw[line width=1pt] (-1.83,5.3) to[bend left=35] (-2.69,3.98);
    \draw[line width=1pt] (-2.7,3.8) to[bend left] (-1.5,5.5);
    \end{scope}
    \draw[densely dashed,thick,seccol] (-8,2) -- (8,2);
    \draw[thin,-Triangle] (9,4) to[bend left] node[above,pos=0] {\shortstack{Orientifold\\plane}}  (8.1,2);
    \node[primcol] at (-2,2) {\small\SnowflakeChevron};
    \node[primcol] at (4,2) {\small\SnowflakeChevron};
    \end{tikzpicture}
    \caption{A representation of the orientifold projection acting on a CY manifold with two double throats. The blue dashed line represent the fixed locus of the projection. The double throats are mapped to each other by the $\mathbb{Z}_2$ symmetry. Snowflakes depict frozen conifolds.}
    \label{fig:orientproj}
\end{figure}

We would like now to bring the attention back to the set of CICY orientifolds constructed in~\cite{Carta:2020ohw}. A subset of them consists of geometries hosting multi-conifolds and therefore thraxions. We immediately stress that these multi-conifolds are \emph{not} the frozen ones discussed in the previous paragraph, as thraxions are defined in the deformed phase. We will explain this point in more details later. In order for a CICY orientifold to allow for the presence of thraxions, two conditions must be satisfied:
\begin{enumerate}
    \item In the double cover, it must be possible to cross a conifold transition locus in a way that preserves the $\ZZ_2$ symmetry that one uses to define the orientifold. As a consequence, the resolved side must have $h^{1,1}_+$ larger than the deformed side.
    \item The set of axions that appears in the resolved side must not be fully projected out by the O$7$-orientifold projection. This means that $h^{1,1}_-$ must also increase in the resolved side. 
\end{enumerate}
The two conditions together imply the following: at the $\mathcal{N}=2$ level, there are two sets of multi-conifolds, each one with the same number of conifold points in it, the same number of homology relations, and the orientifold swaps them. We notice that, despite the multi-conifolds do not lie on the $\ZZ_2$ fixed locus, in principle other sets of conifold singularities can, and generically will, lie on top of the orientifold plane. Therefore, resulting in frozen conifolds. We depict this in Figure~\ref{fig:orientproj}.

Let $\mathcal{M}$ be the set of $319,521$ thraxions transitions. With this we mean couples of resolved and deformed geometries associated with a conifold transition used to define a thraxion. Equivalently, $\mathcal{M}$ is the set of all the possible multi-throat systems that can appear in the CICY orientifold database. We notice that it is possible that the same deformed side of the CY orientifold has more than one multi-throat, and therefore has more than one resolved phase. We define three interesting subsets of $\mathcal{M}$ as follows. First, we consider a set $\mathcal{M}_1$ of multi-throats such that the position and number of O3/O7-planes is the same both in the deformed phase and the resolved one, related by the conifold transition used to define the thraxions themselves. We find that $\mathcal{M}_1$ consists of $11,533$ elements. In addition to this, we further restrict to orientifolds that do not have frozen conifolds. This leaves us with a subset $\mathcal{M}_2\subset\mathcal{M}_1$ of $1,279$ examples satisfying both conditions. Finally, we consider only orientifolds that do admit both O$7$-planes and O$3$-planes. This generates a subset $\mathcal{M}_3\subset\mathcal{M}_2\subset\mathcal{M}_1$ of $57$ examples satisfying all three conditions.

The reason why we restrict to these subsets is the following. For CY orientifolds in $\mathcal{M}\setminus \mathcal{M}_1$ the number and position of orientifold planes varies in a discontinuous way when crossing the transition locus. This means that in the proximity of the transition locus, some O-planes are very close to either merging or splitting. This implies that some extra degrees of freedom become very light in such a region of the moduli space. We leave the study of this very interesting situation for future work. For CY orientifolds in $\mathcal{M}_1\setminus \mathcal{M}_2$ the number and position of O-planes in the two sides of the transition agree, but there are frozen conifold points on at least one of the O7-planes. While these models are in principle viable for phenomenology, the presence of frozen conifolds makes it hard to compute topological quantities needed for writing the low energy effective action, as for example the triple intersection numbers. 
Finally, CY orientifolds in $\mathcal{M}_2\setminus \mathcal{M}_3$ are free of the two possible problems remarked before. However, all orientifolds in this set have either no O7-planes, or no O3-planes. Therefore, using them for phenomenology can be challenging or also completely impossible. 

We compiled a new database, listing all the couples of deformed/resolved CICY orientifolds contained in $\mathcal{M}_2$. This database is explicitly available at this \href{https://www.desy.de/~westphal/orientifold_webpage/cicy_thraxions.html}{link}.
Every element of the database takes the following form:
\begin{equation}
    \left\{ \left\{\text{Resolved CICY info}\right\}, \left\{\text{Deformed CICY info}\right\},\text{\# thraxions}, \text{\# Conifold pts.}\right\}\fstop
    \label{eq:elemdatabase}
\end{equation}
The last two entries are the number of thraxions and the number of conifold points, computed as:
\begin{equation}
    \text{\# thraxions}\coloneqq \left|\Delta h^{1,1}_-\right| \coma \text{\# Conifold pts.}\coloneqq \frac{1}{4}\left|\Delta\chi\right|\coma
\end{equation}
with $\chi$ the Euler number of the two CYs. 
The first two components contain some useful information about the orientifold:
\begin{enumerate}
    \item The number of the CICY following the numeration given in~\cite{Anderson:2017aux}.
    \item $h^{1,1}$ and $h^{2,1}$ of the CICY before the orientifold action.
    \item A configuration matrix of the CICY.
    \item The triple intersection polynomial of the CICY.\footnote{This is computed with respect to a divisor basis given by the pullbacks of the hyperplane classes of the various $\PP^{n_i}\in \mathcal{A}$.}
    \item The number of the orientifold given in~\cite{Carta:2020ohw}.
    \item The rows in the configuration matrix that have to be swapped in order to define the thraxion transition.
    \item $h^{1,1}_-$ and $h^{2,1}_-$ of the CICY orientifold.
    \item The triple intersection polynomial of the CICY orientifold computed as in~\cite{Gao:2013pra}. 
    \item The data relative to the number of O$7$-planes, number of O$3$-planes, number of frozen conifolds on each O$7$-plane, following the notation of~\cite{Carta:2020ohw}.
\end{enumerate}

\begin{table}[!htp]
    \centering
    \begin{tabular}{c|c|c|c|c}
        \multirow{2}{*}{Thraxions} & \multicolumn{3}{c|}{Orientifolds} & \multirow{2}{*}{Total}\\
        \cline{2-4}
          & \shortstack{Both\\O7/O3-planes} & \shortstack{Only\\O7-planes} & \shortstack{Only\\ O3-planes}  &\\
          \hline
        1 & 55 & 140 & 381 & 576 \\
        2 & 1  & 12 & 114 & 127 \\
        3 & 1  & 12  & 41  & 54 \\
        4 & 0   & 0   & 14  & 14 \\
        5 & 0   & 0   & 12  & 12 \\
        6 & 0    & 0   & 9   & 9 \\
        7 & 0    & 0   & 6   & 6 \\
        8 & 0    & 0   & 3   & 3 \\
        9 & 0    & 0   & 2   & 2 \\
        10 & 0   & 0   & 2   & 2 \\
        11 & 0   & 0   & 0   & 0 \\
        12 & 0   & 0   & 0   & 0 \\
        13 & 0   & 0   & 1   & 1 
    \end{tabular}
    \caption{Number of CICY orientifolds in $\mathcal{M}_2$ divided by the number of thraxions and kind of orientifolds they admit.}
    \label{tab:distthrCICYsM2}
\end{table}

\begin{figure}[!htp]
\centering
      \pgfplotstableread{
Label sO7O3 sO7 sO3 topper
1 55 140 381 0
2 1 12 114 0
3 1 12 41 0
4 0 0 14 0
5 0 0 12 0
6 0 0 9 0
7 0 0 6 0
8 0 0 3 0
9 0 0 2 0
10 0 0 2 0
11 0 0 0 0
12 0 0 0 0
13 0 0 1 0
    }\testdata
         \begin{tikzpicture}[scale=0.9]
    \begin{axis}[
        ybar stacked,
        ymin=0,
        ymax=800,
        xtick={data},
        x=8mm,
        legend style={cells={anchor=west}, legend pos=north east},
        reverse legend=false,
        xticklabels from table={\testdata}{Label},
        xticklabel style={text width=2cm,align=center},
        bar width=0.7cm,
        xlabel={Thraxions},
		ylabel={\# CICY orient.},
    ]
    \addplot+[fill=primcol,ybar]
            table [y=sO7O3, meta=Label, x expr=\coordindex]
                {\testdata};
                    \addlegendentry{O7/O3}
        \addplot [fill=seccol,ybar]
            table [y=sO7, meta=Label, x expr=\coordindex]
                {\testdata};
                    \addlegendentry{O7}
        \addplot [fill=tercol,ybar,nodes near coords,point meta=y]
            table [y=sO3, meta=Label, x expr=\coordindex]
                {\testdata};
                    \addlegendentry{O3}
    \end{axis}
    \end{tikzpicture}
    \caption{Number of CICY orientifolds in $\mathcal{M}_2$ for given number of thraxions. The different colors show the presence of both O7/O3-planes, or only O7 or O3-planes. The detailed numbers of CICY divided by the kind of the orientifolds and the number of thraxions are shown in Table~\ref{tab:distthrCICYsM2}.}
    \label{fig:distthrCICYsM2}
\end{figure}
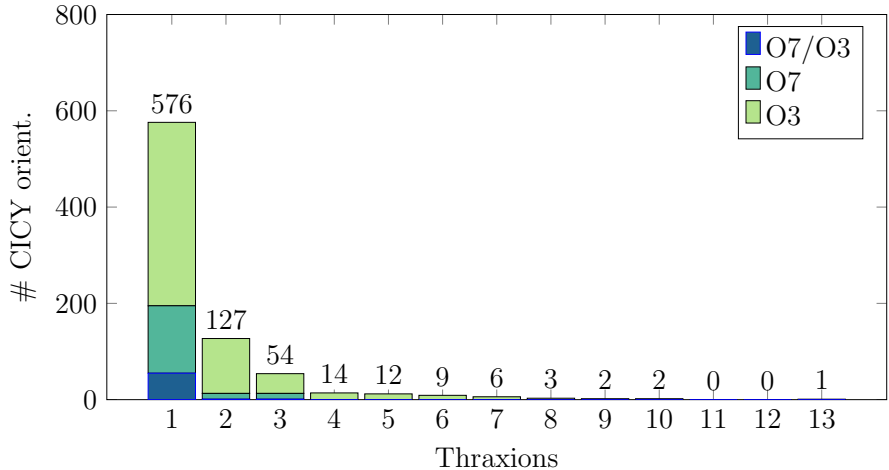

Essentially, by construction, for every multi-throat system there is just one thraxion.\footnote{We stress that this needs not to be the case in general, it is just an artifact of the way in which thraxions transitions were discovered in~\cite{Carta:2020ohw}.} However, it does happen that the same CICY orientifold admits \emph{multiple} multi-throats, therefore allowing for multiple thraxions, still one per multi-throat system. We display in Table~\ref{tab:distthrCICYsM2} and Figure~\ref{fig:distthrCICYsM2} the number of multi-throats (and therefore the number of thraxions) within the database $\mathcal{M}_2$ discussed above.

\begin{table}[!htp]
    \centering
    \begin{tabular}{c|c|c|c|c}
        \multirow{2}{*}{Thraxions} & \multicolumn{3}{c|}{Orientifolds} & \multirow{2}{*}{Total}\\
        \cline{2-4}
          & \shortstack{Both\\O7/O3-planes} & \shortstack{Only\\O7-planes} & \shortstack{Only\\ O3-planes}  &\\
          \hline
        1 & 7117 & 900 & 381 & 8398 \\
        2 & 721  & 136 & 114 & 971 \\
        3 & 119  & 29  & 41  & 189 \\
        4 & 15   & 4   & 14  & 33 \\
        5 & 25   & 6   & 12  & 43 \\
        6 & 8    & 4   & 9   & 21 \\
        7 & 0    & 0   & 6   & 6 \\
        8 & 0    & 0   & 3   & 3 \\
        9 & 4    & 0   & 2   & 6 \\
        10 & 0   & 0   & 2   & 2 \\
        11 & 0   & 0   & 0   & 0 \\
        12 & 0   & 0   & 0   & 0 \\
        13 & 0   & 0   & 1   & 1 
    \end{tabular}
    \caption{Number of CICY orientifolds in $\mathcal{M}_1$ divided by the number of thraxions and kind of orientifolds they admit.}
    \label{tab:distthrCICYsM1}
\end{table}

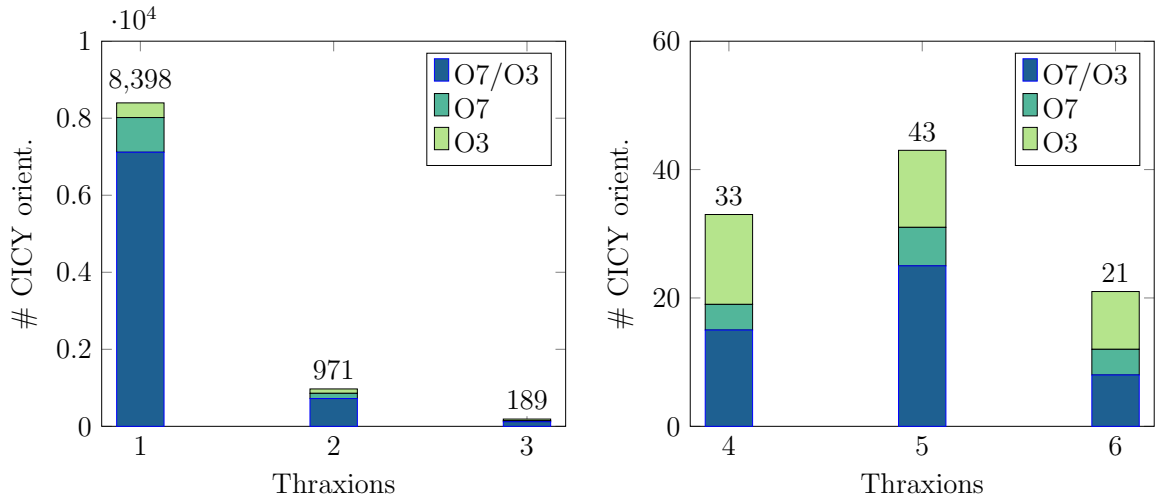
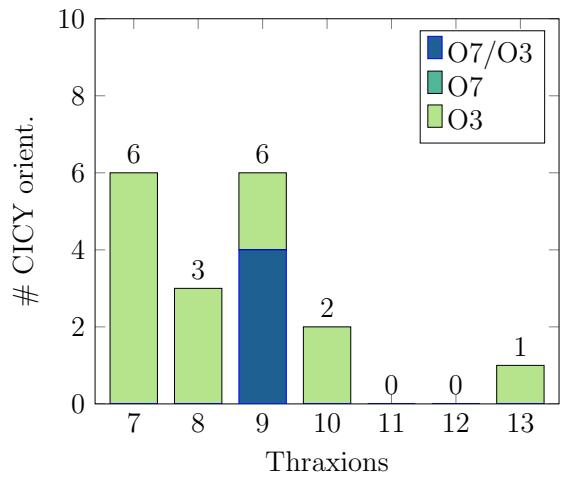
\begin{figure}[!htp]
    \centering
    \begin{subfigure}[t]{0.49\textwidth}
\centering
      \pgfplotstableread{
Label sO7O3 sO7 sO3 topper
1 7117 900 381 0
2 721 136 114 0
3 119 29 41 0
    }\testdata
         \begin{tikzpicture}[scale=0.9]

    \begin{axis}[
        ybar stacked,
        ymin=0,
        ymax=10000,
        xtick={data},
        legend style={cells={anchor=west}, legend pos=north east},
        reverse legend=false,
        xticklabels from table={\testdata}{Label},
        xticklabel style={text width=2cm,align=center},
        bar width=0.7cm,
        xlabel={Thraxions},
		ylabel={\# CICY orient.},
    ]
    \addplot+[fill=primcol,ybar]
            table [y=sO7O3, meta=Label, x expr=\coordindex]
                {\testdata};
                    \addlegendentry{O7/O3}
        \addplot [fill=seccol,ybar]
            table [y=sO7, meta=Label, x expr=\coordindex]
                {\testdata};
                    \addlegendentry{O7}
        \addplot [fill=tercol,ybar,nodes near coords,point meta=y]
            table [y=sO3, meta=Label, x expr=\coordindex]
                {\testdata};
                    \addlegendentry{O3}
    \end{axis}
    \end{tikzpicture}
\caption{Number of CICY orientifolds that admit up to $3$ thraxions.}
  \end{subfigure}\hfill
     \begin{subfigure}[t]{0.49\textwidth}
\centering
      \pgfplotstableread{
Label sO7O3 sO7 sO3 topper
4 15 4 14 0
5 25 6 12 0
6 8 4 9 0
    }\testdata
         \begin{tikzpicture}[scale=0.9]

    \begin{axis}[
        ybar stacked,
        ymin=0,
        ymax=60,
        xtick={data},
        legend style={cells={anchor=west}, legend pos=north east},
        reverse legend=false,
        xticklabels from table={\testdata}{Label},
        xticklabel style={text width=2cm,align=center},
        bar width=0.7cm,
        xlabel={Thraxions},
		ylabel={\# CICY orient.},
    ]
    \addplot+[fill=primcol,ybar]
            table [y=sO7O3, meta=Label, x expr=\coordindex]
                {\testdata};
                    \addlegendentry{O7/O3}
        \addplot [fill=seccol,ybar]
            table [y=sO7, meta=Label, x expr=\coordindex]
                {\testdata};
                    \addlegendentry{O7}
        \addplot [fill=tercol,ybar,nodes near coords,point meta=y]
            table [y=sO3, meta=Label, x expr=\coordindex]
                {\testdata};
                    \addlegendentry{O3}
    \end{axis}
    \end{tikzpicture}
\caption{Number of CICY orientifolds that admit from $4$ to $6$ thraxions.}
  \end{subfigure}\\
       \begin{subfigure}[t]{0.49\textwidth}
\centering
      \pgfplotstableread{
Label sO7O3 sO7 sO3 topper
7 0 0 6 0
8 0 0 3 0
9 4 0 2 0
10 0 0 2 0
11 0 0 0 0
12 0 0 0 0
13 0 0 1 0
    }\testdata
         \begin{tikzpicture}[scale=0.9]

    \begin{axis}[
        ybar stacked,
        ymin=0,
        ymax=10,
        xtick={data},
        legend style={cells={anchor=west}, legend pos=north east},
        reverse legend=false, 
        xticklabels from table={\testdata}{Label},
        xticklabel style={text width=2cm,align=center},
      bar width=0.7cm,
      xlabel={Thraxions},
		ylabel={\# CICY orient.},
    ]
    \addplot+[fill=primcol,ybar]
            table [y=sO7O3, meta=Label, x expr=\coordindex]
                {\testdata};
                    \addlegendentry{O7/O3}
        \addplot [fill=seccol,ybar]
            table [y=sO7, meta=Label, x expr=\coordindex]
                {\testdata};
                    \addlegendentry{O7}
        \addplot [fill=tercol,ybar,nodes near coords,point meta=y]
            table [y=sO3, meta=Label, x expr=\coordindex]
                {\testdata};
                    \addlegendentry{O3}
    \end{axis}
    \end{tikzpicture}
\caption{Number of CICY orientifolds that admit from $7$ to $13$ thraxions.}
  \end{subfigure}
    \caption{Number of CICY orientifolds in $\mathcal{M}_1$ for given number of thraxions. The different colors show the presence of both O7/O3-planes, or only O7 or O3-planes. The detailed numbers of CICY divided by the kind of the orientifolds and the number of thraxions are shown in Table~\ref{tab:distthrCICYsM1}.}
    \label{fig:distthrCICYsM1}
\end{figure}

If we instead consider the set $\mathcal{M}_1$, we find a much larger number of orientifolds supporting thraxions. We report this in Table~\ref{tab:distthrCICYsM1} and Figure~\ref{fig:distthrCICYsM1}. 

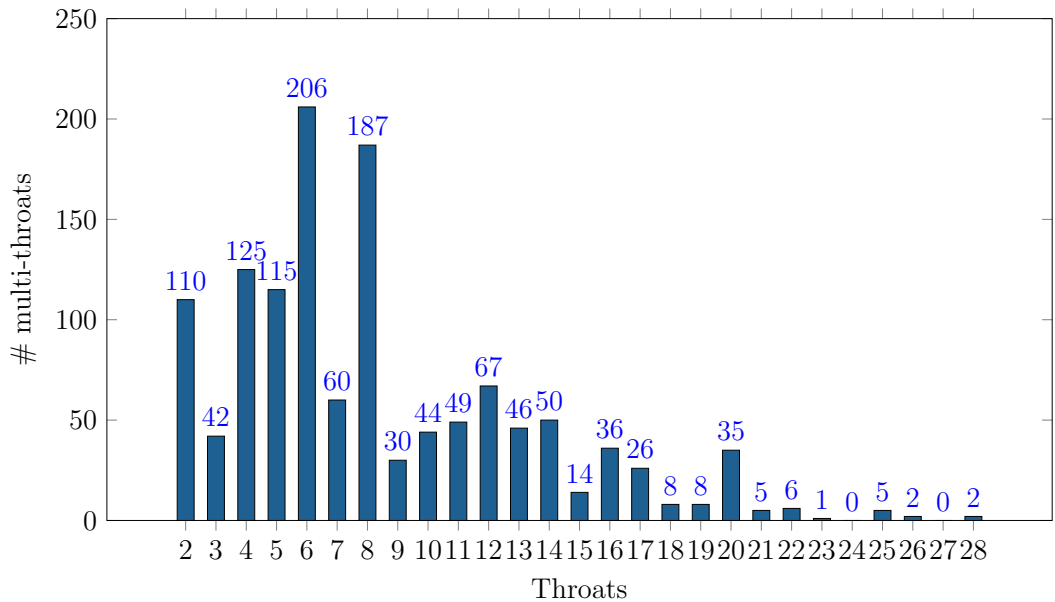
\begin{figure}
 \centering
      \pgfplotstableread{
Label scp topper
2 110 0
3 42 0
4 125 0
5 115 0
6 206 0
7 60 0
8 187 0
9 30 0
10 44 0
11 49 0
12 67 0
13 46 0
14 50 0
15 14 0
16 36 0
17 26 0
18 8 0
19 8 0
20 35 0
21 5 0
22 6 0
23 1 0
24 0 0
25 5 0
26 2 0
27 0 0
28 2 0
    }\testdata
         \begin{tikzpicture}[scale=0.9]
    \begin{axis}[
        ybar,
        ymin=0,
        ymax=250,
        xtick={data},
        width=\textwidth,
        height=9cm,
      xticklabels from table={\testdata}{Label},
        xticklabel style={text width=2cm,align=center},
      bar width=2.5mm,
      xlabel={Throats},
		ylabel={\# multi-throats},
    ]
    \addplot+[fill=primcol,ybar,nodes near coords,point meta=y,draw=black]
            table [y=scp, meta=Label, x expr=\coordindex]
                {\testdata};
    \end{axis}
    \end{tikzpicture}
  \caption{Number of multi-throats in $\mathcal{M}_2$ that admit from $2$ to $28$ throats.}
  \label{fig:numconM3}
\end{figure}

We would like to comment now about the following fact. Since in every multi-throat system there is a single homology relation giving rise to a single thraxion, when we study the moduli stabilization problem we are in the situation described in Section~\ref{sec:structurePotdt}. Therefore, it is possible to argue that with a certain amount of tuning of the fluxes, the thraxion potential does not receive order $\mathcal{O}(\epsilon)$ contributions from the stabilization of the K\"ahler moduli. Such needed tuning involves a democratic distribution of the fluxes in the throats and as a result the thraxion mass is still six-times-warped suppressed. However, if the number of throats in a given multi-throat system is equal to $2$, the tuning of the fluxes is minimal. For this reason, in Figure~\ref{fig:numconM3} we show the multi-throats in $\mathcal{M}_2$ for a given number of throats. In the database we provide, there are, then, $110$ multi-throats that have only $2$ conifold points. We leave for further study the question of which exact flux choice must be made so that the discussion in Section~\ref{sec:structurePotdt} can be realized. 

\FloatBarrier

\section{Conclusions}
In this work, we performed a detailed study about thraxions beyond the flux compactification level. 
Specifically, we analyze $4$d, $\mathcal{N}=1$ effective supergravity in presence of thraxions and how the results derived in~\cite{Hebecker:2018yxs} change when also the K\"ahler moduli sector is taken into account. 

In the first part of this work, 
we study the backreaction of a non-vanishing thraxion VEV on the internal space geometry. By analyzing $\SU(3)$-structure's torsion classes, we find that the CY condition is broken due to the breakdown of the ISD condition of the $G_3$ flux. This leaves us with just a complex manifold. The amount of the breaking is related qualitatively to the value of the thraxion VEV. In turn, if the CY condition is broken already at the KK scale, the use of the 4d supergravity approximation  in order to describe the $4$d effective theory could be questionable. However, we argue for a sufficiently small thraxion VEV or a decoupling of the thraxion dynamics coming from the high warping, in such a way that the manifold is still (almost) CY. Hence, we can be entitled to use the effective supergravity action and include the K\"ahler moduli stabilization.

The second part of this work aims at studying the relation of thraxions and K\"ahler moduli in the presence of perturbative and non-perturbative corrections to the K\"ahler potential and the superpotential. We find that in general K\"ahler moduli stabilization spoils the high suppression of the thraxion mass coming from the sixth power of the warp factor. The no-scale breaking terms induce additional contributions to the potential which are proportional to the warp factor cubed only, hence lifting the thraxion mass. However, the thraxion is still the lightest particle in the spectrum, and the spectrum is still effectively gapped. 

One may ask what are the consequences of this new thraxion behavior on the axionic version of the WGC~\cite{Arkani-Hamed:2006emk}. In~\cite{Hebecker:2018yxs}, it was found a parametric violation of the lattice WGC while the sub-lattice WGC~\cite{Heidenreich:2015nta,Heidenreich:2016aqi,Andriolo:2018lvp} was still satisfied but with a parametrically coarse sub-lattice. The new scaling of the thraxion mass still provides a violation of the lattice WGC but milder by a factor of $2$, resulting in a less coarse sub-lattice needed to satisfy the sublattice WGC. 
The study of the validity of the EFT whenever the thraxion stabilizes to a VEV different from zero is an important task for future work. As explained in Section~\ref{sec:ThraxionsTorclass}, in this situation, the CY condition is broken. It would be interesting to see for how long the EFT is still valid before new light states must be integrated-in. Indeed, the violation of the lattice-WGC might be a symptom that some new objects must be considered in order to fully describe the physics of thraxions.

In addition, there may be implications from the finite D3-brane charge tadpole of any given type IIB CY compactification. While we leave a full discussion of this for future work, we observe that: i) metastability of a dS uplift from a single anti-D3-brane implies a lower bound $M\gtrsim  {\cal O}(10)$ on the RR 3-form flux on the throat $\mathcal{A}$-cycle. ii) Fixing the thraxion mass scale for some physical application on the other hand fixes the warp factor and thus $K/M\gg 1$. This entails the D3-brane charge tadpole placing an upper bound $M<\sqrt{Q_{D3} (K/M)^{-1}}$, which may get tightened by replacing $Q_{D3}$ by $Q_{D3}^{\text{eff.}}< Q_{D3}$ due to flux stabilization of the c.s. moduli eating up a part of $Q_{D3}$~\cite{Kachru:2002gs,Bena:2018fqc,Bena:2020xrh}. Depending on the size of the available tadpole and the number of non-zero fluxes needed to freeze the c.s. moduli, this may effectively limit the achievable warp factor suppression of the thraxion mass scale and/or the existence of  meta-stable anti-D3 uplifts.

 After our generic discussion, we work out the conditions sufficient to obtain the original scaling of the thraxion mass with respect to the warp factor. Such cases require the presence of only one thraxion in each multi-throat system, a democratic distribution of fluxes in the throats and a particular homology relation among the interpolating cycles. These requirements need a certain amount of tuning, unless the multi-throats are double-throats. Therefore, which of the two cases is more prevalent in the string landscape will depend significantly on the relative frequencies of double versus higher multi-throats. It is interesting to note here that in the database of the CICY orientifolds built in~\cite{Carta:2020ohw}, we always have one thraxion per multi-throat system. Whether this is the case also for other CY sets as e.g. the Kreuzer-Skarke database~\cite{Kreuzer:2000xy} is an important question that we leave to further study. 

We then specialize the discussion of no-scale breaking effects to the specific examples of LVS and KKLT moduli stabilization approaches. In particular, in LVS the exponentially large values of the CY volume allow us to suppress the cross terms which otherwise would be responsible for lifting the mass. This suppression turns out to be effective only in scenarios where the scalar potential scale is sufficiently high. This in particular includes the interesting case of high-scale inflation.

Finally, starting from the database of CICY orientifolds presented in~\cite{Carta:2020ohw}, we build a new specific database for thraxions that one can find at this \href{https://www.desy.de/~westphal/orientifold_webpage/cicy_thraxions.html}{link}. Our database aims at showing that there exists the possibility to make phenomenologically viable models with thraxions, and how frequent they are within a given CY orientifold. If we allow for the presence of frozen conifolds in the sense of~\cite{Carta:2020ohw}, we have plenty of possibilities to do model building with thraxions. However, the presence of frozen conifolds prevents the computation of essential quantities, which makes the construction of explicit models challenging at the moment. As frozen conifolds are generically present in CICY orientifolds, restricting to the cases in which they are not present naturally narrows the thraxion landscape. Still, we find 57 CICY orientifolds allowing for the presence of thraxions and both O3 and O7 orientifold planes, and the absence of frozen conifolds.  

Given the obvious phenomenological possibilities of thraxion models, as well as their intrinsic theoretical interest, we hope to come back in the near future to the many interesting open questions we raised in this work.

\section*{Acknowledgments}

We would like to thank Vicente Cortés, Matteo Gallone, Veronica Guidetti, Arthur Hebecker, Jakob Moritz, Enrico Parisini for useful discussions. F.C. is supported by STFC consolidated grant ST/T000708/1. A.M. received funding from ``la Caixa" Foundation (ID 100010434) with fellowship code LCF/BQ/IN18/11660045 and from the European Union’s Horizon 2020 research and innovation programme under the Marie Sk\l odowska-Curie grant agreement No. 713673 until September 2021. The work of A. M. is supported in part by Deutsche Forschungsgemeinschaft under Germany's Excellence Strategy EXC 2121 Quantum Universe 390833306. N.R. is supported by the Deutsche Forschungsgemeinschaft under Germany's Excellence Strategy - EXC 2121 ``Quantum Universe'' - 390833306. A.W. is supported by the ERC Consolidator Grant STRINGFLATION under the HORIZON 2020 grant agreement no. 647995.

\appendix

\section{Details on Moduli Stabilization and Explicit Examples}
\label{app:modstabdetails}
In this Appendix we provide further examples of KKLT and LVS moduli stabilization in presence of a thraxion. We explore setups which are the most relevant extensions of the ones shown in Section~\ref{sec:ModuliStabilization}, by allowing for the presence of more K\"ahler moduli and one thraxions. Our main point here is to show that, in the special cases described in Section~\ref{sec:structurePotdt} (namely, in the cases where there is only one thraxion per multi-throats and where we allow for a certain choice of fluxes), the minimization of the $C_4$ axions always removes the terms in the scalar potential which are linear in $\epsilon$.

\subsection{LVS}\label{app:LVS}
 In this Appendix, we extend the toy model presented in Section~\ref{sec:MS-LVS} to the case of two K\"ahler moduli and one thraxion with both nontrivial triple intersection numbers. In Section~\ref{sec:MS-LVS} we considered one thraxion coupling to the blow-up modulus only. Here, we allow for the coupling with both moduli, i.e. we consider both triple intersection numbers $\kappa_{b--}\equiv\kappa_b$ and $\kappa_{s--}\equiv\kappa_s$ to be nontrivial. We work with a CY parametrized by a Swiss-cheese volume as
 \begin{equation}
 		\mathcal{V}=\left(T_b+\bar{T}_b-\frac{g_s}{4}\kappa_{b}\left(G-\bar{G}\right)^2\right)^{3/2} - \left(T_s+\bar{T}_s-\frac{g_s}{4}\kappa_{s}\left(G-\bar{G}\right)^2\right)^{3/2}
 		\fstop
 \end{equation}
 The superpotential of Eq.~\eqref{eq:Wthraxnonpert} reads
  \begin{equation}
 	W(G,T_s)=W_0+ \epsilon \left(1-\cos\left(G/M\right)\right) +A_s e^{-a_s T_s}\coma
 \end{equation}
 where the leading non-perturbative correction comes from the blow-up modulus. The positivity of the K\"ahler matrix requires $\kappa_b<0$, while $\kappa_s$ is unconstrained.  
 The $b$ field stabilizes at vanishing VEV.
Let us display here only $V_{\text{thr}}$, as the part of the potential independent on the thraxion scales as the classical LVS potential. Then, the potential for the thraxion $c$ reads
 \begin{equation}
 	\begin{split}
 V_{\text{thr}}= &-\frac{3\,\epsilon^2\hat{\xi} }{\mathcal{V}^3}\sin\left(\frac{c}{2M}\right)^4+
 \frac{2\, \epsilon^2\, \tau_s \kappa_{s}^2}{3 g_s M^2 \mathcal{V}^3 \kappa_{b}^6}\sin\!\left(\frac{c}{M}\right)^2\! \left(4\sqrt{2}\tau_s^{3/2}\!\left(\kappa_{s}^3-\kappa_b^3\right)-\hat{\xi}\kappa_{b}^3\right)+\\
 &+\frac{\epsilon^2 \sin\left(\frac{c}{M}\right)^2}{3 \,g_s M^2\, \mathcal{V}^2\, \kappa_{b}^3}\left(2\kappa_b^2\mathcal{V}^{2/3}+2\sqrt{2\tau_s}\kappa_b\kappa_s^4\mathcal{V}^{1/3}+4\tau_s\kappa_s^2\right)+\\
 &+\frac{\epsilon^2 \sin\left(\frac{c}{M}\right)^2}{9\, g_s M^2\, \mathcal{V}^{7/3}\, \kappa_{b}^4}\left(4\sqrt{2}\tau_s^{3/2}\left(3\kappa_{s}^3-\kappa_b^3\right)-3\hat{\xi}\kappa_{b}^3\right)+\\
 &+\frac{\epsilon^2 \sin\left(\frac{c}{M}\right)^2}{9 \,g_s M^2\, \mathcal{V}^{8/3}\, \kappa_{b}^5}\,\kappa_s\sqrt{\tau_s}\,\left(8\tau_s^{3/2}\left(3\kappa_{s}^3-2\kappa_b^3\right)-3\sqrt{2}\hat{\xi}\kappa_{b}^3\right)+
     \\
 		&+\frac{4 i\, \epsilon\, a_s A_s \tau_s e^{-a_s \tau_s}}{\mathcal{V}^2}\left(1-\cos\!\left(\frac{c}{M}\right)\right)\sin\!\left(a_s \theta_s\right)    \fstop
 	\end{split}
 \end{equation}

\subsection{KKLT}

Consider the case in which we have one thraxion intersecting with one K\"ahler modulus, but in a setup with $h^{1,1}_+=2$. Suppose we can write the K\"ahler potential as
\begin{equation}
	K_{\text{thr}}=-2 \log\left(\left(T_1+\bar{T}_1\right)^{3/2}+\left(T_2+\bar{T}_2-\frac{g_s}{4}\kappa_{2 --}\left(G-\bar{G}\right)\right)^{3/2}\right)\fstop
 \end{equation}
Then, by using the superpotential in Eq.~\eqref{eq:Wthraxnonpert}, we can compute the scalar potential for the K\"ahler moduli, the $C_4$ axions and the thraxion. For ease of exposition, we display only the terms dependent on $c$ after setting $b$ to its VEV, namely
\begin{equation}
    \begin{split}
    V_{\text{thr}}=
   & 
   \,\frac{\epsilon^2\sin\!\left(\frac{c}{M}\right)^2}{6 g_s M^2\kappa_{2--}\sqrt{\tau_2}\left(\tau_1^{3/2}+\tau_2^{3/2}\right)}+\\
   &+\frac{i\,\epsilon\sin\!\left(\frac{c}{2M}\right)^2}{\left(\tau_1^{3/2}+\tau_2^{3/2}\right)^2}\left(a_1 A_1 e^{-a_1 \tau_1} \sin\!\left(a_1\theta_1\right)+a_2 A_2 e^{-a_2 \tau_2} \sin\!\left(a_2\theta_2\right)\right)
   \fstop
    \end{split}
\end{equation}
We see therefore that there is a repetition of the case with one modulus only of Section~\ref{sec:structurePotdt}, where this time the cross term vanishes when both the $C_4$ axions $\theta_1$, $\theta_2$ are set to their vanishing VEVs.

\section{The Mass of the KKLT AdS Tachyon}
\label{app:KKLT-repth}
In this section, we consider a KKLT scenario with only one K\"ahler modulus. We will not take into account the potential for the thraxions, although they will enter as dynamical fields.  We are going to check that the masses obtained for the fields $b$ and $c$ in the KKLT AdS vacuum are consistent with results from representation theory of the AdS isometry group $\SO(3,2)$. 

Let us first describe in general how the masses for the moduli are obtained, in a generic supergravity model, following~\cite[Section 3.1]{Conlon:2007gk}. Suppose that we have found the minimum of a potential for the moduli $\phi_i$. Around the minimum we can write the moduli as
\begin{equation}
    \phi_i = \langle \phi_i \rangle + \delta \phi_i = \langle \phi_i \rangle + \varphi_i\fstop
\end{equation}
The Lagrangian for the fluctuations $\varphi_i$ is 
\begin{equation}
    \mathcal{L}=K_{i\bar{j}}\de_\mu \varphi^i \de^\mu \varphi^j -\left(M^2\right)_{ij}\varphi^i\varphi^j -V_0 -\mathcal{O}\left(\varphi^3\right)\fstop
    \label{eq:startingpoint}
\end{equation}
We stress that $K_{i\bar{j}}$ and $(M^2)_{ij}$ are computed in the values of the moduli at the minimum.\\
In our specific setup, both $K_{i\bar{j}}$ and $(M^2)_{ij}$ are diagonal matrices. It is then possible to define 
\begin{equation}
    \Phi_i = \sqrt{2} K_{i\bar{i}}\varphi_i \Longrightarrow \varphi_i = \frac{1}{\sqrt{2}} (K^{-1})_{i\bar{i}}\Phi_i\fstop
    \label{eq:fieldstransform}
\end{equation}
Using this definition, we obtain the following Lagrangian for the canonically normalized fields $\Phi_i$:
\begin{equation}
    \mathcal{L}=\frac{1}{2}\de_\mu \Phi_i \de^\mu \Phi_i -\frac{1}{2}(M^2)_{ii}(K^{-1})_{ii} \Phi_i \Phi_i -V_0 - \mathcal{O}(\Phi^2)\fstop
\end{equation}
For a more general case, we refer to~\cite{Conlon:2007gk}. The physical masses of the fields are given by $(M^2)_{ii}(K^{-1})_{ii}$. 

Consider the K\"ahler potential in the case of a single K\"ahler modulus $T$ and thraxion $G$ and the superpotential in~\eqref{eq:1dtsuperpot} with $\epsilon=0$.\footnote{We are then considering only the non-perturbative superpotential for the K\"ahler moduli.}
The corresponding $\phi_i$ in Eq.~\eqref{eq:startingpoint} are then $\{\tau,\theta,b,c\}$. Minimizing the potential \eqref{eq:VSUGRA}, we find that the values of the fields at the minimum are
\begin{equation}
   \langle \theta \rangle =0 \coma \langle b \rangle =0 \coma
    \label{eq:VEVmoduli}
\end{equation}
while $c$ is left unstabilized. The K\"ahler modulus $\tau$ will also be stabilized at some given value $\langle\tau\rangle$. 
Computing $(M^2)_{ii}(K^{-1})_{ii}$ at the minimum of the potential, the value for the $b$ mass is given by
\begin{equation}
    m_b^2=-\frac{1}{g_s^2}\frac{a^2 |W_0|^2}{ \langle\tau\rangle  (2 a \langle\tau\rangle +3)^2}\fstop
    \label{eq:mbKKLT}
\end{equation}
Being all the quantities positive definite, clearly $b$ has a negative mass. 

Let us check that this result is consistent with representation theory. The fields $\{b,c\}$ belong to the same massive Wess-Zumino supermultiplet. The field content of this multiplet is given by a real scalar $0^+$, a real pseudoscalar $0^-$ and a spin-$1/2$ fermion, subject to the relation their masses are~\cite{Dusedau:1988nc,deWit:1999ui,Gunara2003:aaa}
\begin{equation}
    m_{0^+}^2=m^2+m\ell -2\ell^2 \coma m_{0^-}=m^2-m\ell -2\ell^2 \,\,\text{ and }\,\, m_{1/2}^2=m^2 \coma
    \label{eq:massWessZumino}
\end{equation} where $\ell$ is proportional to the cosmological constant. Notice that, compared to SUSY QFTs in Minkowski space, the operator $M^2=P_{\mu}P^{\mu}$ is no longer commuting with the supercharges, and therefore different fields in the same multiplet will have different masses. In our case the pseudoscalar field is the $c$ axion, the real scalar field is the $b$ axion, and $\zeta$ is the fermion.

Defining $V_0$ to be the potential at the minimum, we have
\begin{equation}
    \ell=\sqrt{\frac{V_0}{-3}} =\frac{1}{\sqrt{2}}  \frac{a |W_0|}{ \langle\tau\rangle^{1/2} (3 + 2 a \langle\tau\rangle)}\fstop
\end{equation}
In our setup, the field $c$ has zero mass, as it is a flat direction of the potential. Therefore, by identifying it with the pseudoscalar in the Wess-Zumino multiplet, we can set to zero the second equation in Eq.~\eqref{eq:massWessZumino}. Solving this in $m$, we find that $m_{0^-}$ is zero when 
\begin{equation}
    m= -\ell \,\text{ or }\, m=2\ell\fstop
\end{equation}
Choosing the first possibility and substituting in the mass for $m_{0^+}$, we get
\begin{equation}\label{eq:represtheorymass}
    m_{0^+}=-\frac{a^2 |W_0|^2}{ \langle\tau\rangle  (2 a \langle\tau\rangle +3)^2}\coma
\end{equation}
that exactly matches the computation from the Hessian in~\eqref{eq:mbKKLT}.\footnote{Up to a rescaling of the fields to add $g_s$.} 
Finally, we notice that the negative mass of $b$ is still compatible with the BF bound; indeed, it saturates it.

\newpage

\bibliographystyle{JHEP}
\bibliography{mybib}

\end{document}